\documentclass[journal]{IEEEtran}

\usepackage{amsmath,amssymb,amsthm,amsfonts,mathtools} 
\usepackage{dsfont,eucal,bbm,bm,nicefrac} 
\usepackage{graphicx,float,subcaption,booktabs} 
	\graphicspath{{./figures/}}
\usepackage{algorithm,algorithmic} 
\usepackage{hyperref} 
\usepackage{tikz,xcolor,soul} 
\usepackage{cite}
\usepackage{textcomp}
\usepackage{hyperref}
\hypersetup{hidelinks}

\newtheorem{theorem}{Theorem}
\newtheorem{corollary}{Corollary}[theorem]

\newtheorem{remark}{Remark}

\newtheorem{definition}{Definition}

\newcommand{\pbra}[1]{\ensuremath{\left( #1\right)}}
\newcommand{\sbra}[1]{\ensuremath{\left[ #1\right]}}
\newcommand{\cbra}[1]{\ensuremath{\left\{ #1\right\}}}

\newcommand{\E}[1]{\ensuremath{\mathbb{E}\left[ #1\right]}}

\newcommand{\Ep}[2]{\ensuremath{\mathbb{E}_{#1}\left[ #2\right]}}

\DeclareMathOperator*{\argmax}{arg\,max}
\DeclareMathOperator*{\argmin}{arg\,min}

\DeclareMathOperator*{\minimize}{min.}
\DeclareMathOperator*{\maximize}{max.}

\begin{document}

\title{
Risk-Sensitive Reinforcement Learning \\with Exponential Criteria}

\author{Erfaun Noorani$^*$, Christos N. Mavridis$^\dagger$, and John S. Baras$^*$ 
%
\thanks{$^*$%
Department of Electrical and Computer Engineering and 
the Institute for Systems Research, 
University of Maryland, College Park, USA.
{\tt\small emails: \{noorani,baras\}@umd.edu}.}%
\thanks{$^\dagger$%
Division of Decision and Control Systems, 
School of Electrical Engineering and Computer Science,
KTH Royal Institute of Technology, Stockholm.
{\tt\small email: mavridis@kth.se}.}%
\thanks{%
This work was supported in part by US Office of Naval Research (ONR) Grant No. N00014-17-1-2622, by US NSF Grant ECCS2127605, by the Clark Foundation Distinguished Fellowship, and by the University of Maryland Graduate School Ann G. Wylie Fellowship.
}%
}

\maketitle
 \thispagestyle{empty}
\pagestyle{empty}

\begin{abstract}
While reinforcement learning has shown experimental success in a number of 
applications, it is known to be sensitive 
to noise and perturbations in the parameters of the system, 
leading to high variability in the total reward amongst different episodes on slightly 
different environments.
To introduce robustness, as well as sample efficiency, 
risk-sensitive reinforcement learning methods are being thoroughly studied.
In this work, 
we provide a definition of robust reinforcement learning policies and 
formulate a risk-sensitive reinforcement learning problem to approximate them,
by solving an optimization problem with respect to a modified objective based on 
exponential criteria.
In particular, 
we study a model-free risk-sensitive variation of the widely-used 
Monte Carlo Policy Gradient algorithm, and introduce a novel 
risk-sensitive online Actor-Critic algorithm based on 
solving a multiplicative Bellman equation using stochastic approximation updates.
Analytical results suggest that the use of exponential criteria 
generalizes commonly used ad-hoc regularization approaches, 
improves sample efficiency, and 
introduces robustness with respect to perturbations in the model parameters and 
the environment.
The implementation, performance, and robustness properties of the proposed methods
are evaluated in simulated experiments.
\end{abstract}


\section{Introduction}
\label{Sec:Introduction}

\begin{figure}[t] 
    \centering
    \includegraphics[trim=10 10 10 5,clip,height=0.175\textwidth]{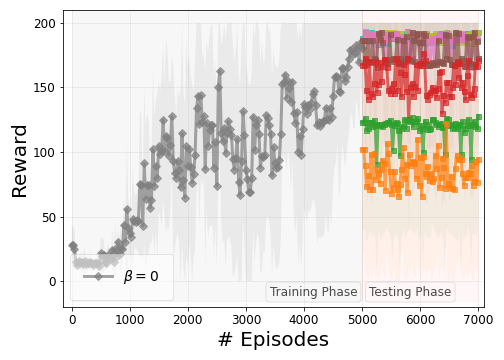}
    \includegraphics[trim=35 10 10 5,clip,height=0.175\textwidth]{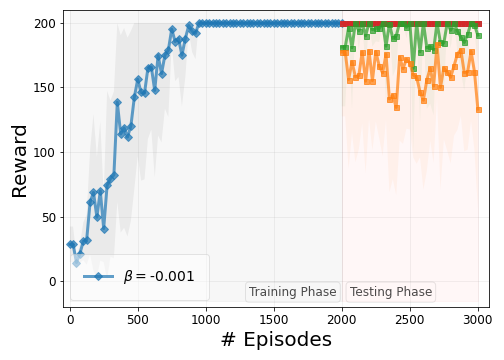}
    \hspace{-1.1cm}
    \includegraphics[trim=0 -30 0 0,clip,width=0.05\textwidth]{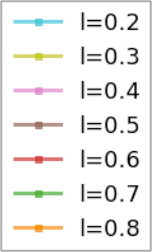}
    \caption{Generalization performance with respect to perturbations in the model parameters. Risk-neutral (left) and risk-sensitive (right) actor-critic reinforcement learning algorithms trained in the Cart-Pole environment with pole length $l=0.5$
    are tested for different pole length values $l\in\sbra{0.2,0.8}$.
    Average reward and $90\%$ confidence intervals over a running window of $10$ episodes are depicted.}
    \label{fig:cartpole-generalization}
    \vspace{-1em}
\end{figure}

In stochastic decision systems, where uncertainty leads to risk (variability) in a desired performance metric, 
solving a stochastic optimal control task 
(viz., reinforcement learning applications) 
by optimizing a risk-neutral objective, often leads to 
control policies that might perform poorly, especially in real-world applications. 
This is due to the fact that risk-neutral objectives typically consist of a long-run 
expectation of the desired metric (average performance)
which have been shown to be non-robust to noise and 
model uncertainties \cite{sutton2018reinforcement}. 
This phenomenon is observed in widely-used Reinforcement Learning (RL) algorithms, such as Actor-Critic methods, 
which are often unable to maintain their performance under slight variations in the environment at the testing time. 
Figure \ref{fig:cartpole-generalization} shows the training and testing performance of an Actor-Critic agent in an inverted pendulum problem (see Section \ref{sSec:cartpole}) 
with perturbed model parameters. 
While training is conducted with a given pole length, the performance of the trained agent is evaluated in a set of environments with different pole lengths. 
It is clear that, in the risk-neutral case, the change in the pole length results in significant performance degradation. 
To mitigate such issues, risk-sensitive RL investigates alternative optimization approaches, 
by incorporating constraints and alternative objective functions 
to induce robustness with respect to variations and uncertainties of the environment \cite{moos2022robust,osogami2012robustness,Noorani2021REINFORCE}. 
%

\subsection*{Related work}
Robustness has been studied extensively in optimization and optimal control \cite{hansen2011robustness}.
In reinforcement learning problems, where uncertainties in the system 
demand that distributional information is taken into account, 
robustness is associated with a stochastic optimization problem of the form:
\begin{equation*}
    \max_{\pi\in \Pi}. \ \inf_{\rho\in \Psi} \ \Ep{x\sim \pi, \zeta\sim \rho}{R(x,\zeta)},
\end{equation*}
where $x\in X$ are the design parameters with distribution $\pi\in \Pi$, 
$\zeta\in Z$ is a random vector 
with distribution $\rho\in \Psi$ representing uncertain system parameters, and $R:X\times Z\rightarrow [ 0,\infty)$ is an objective (reward) function to be maximized.
Here the system's sensitivity to maximum uncertainty (e.g., noise, disturbances) is maximized
\cite{scarf1957min}.
This problem is closely related to mini-max games \cite{
james1995robust}.

A number of risk-sensitive reinforcement learning approaches have been studied in recent years;
from constructing constraint stochastic optimization problems
\cite{paternain2019constrained,delage2010percentile,abdullah2019wasserstein} or 
approximately solving mini-max optimization problems \cite{tamar2013scaling},
to investigating different statistical measures of the objective function \cite{fu2018risk}.
%
The latter approach often provides benefits to algorithmic implementations, 
since the computational problems associated with constraint optimization and 
the problems associated with the existence of multiple Nash equilibria, are avoided.
In particular, the algorithms in \cite{chow2014algorithms,prashanth2014policy,tamar2015risk,pinto2017robust} use the conditional value at risk for policy search and the algorithms in \cite{pan2019risk,dotan2012policy,liu2018r2pg,eysenbach2021maximum} use variance as the desired risk measure. 
%
%
%

Although these are ad-hoc approaches developed
by experimental observations, 
there is a duality connection between KL- and entropy-regularized objectives and 
entropic risk measures 
\cite{borkar2010learning,osogami2012robustness,nass2019entropic,Noorani2021risk}, 
associated with exponential criteria of the form:
%
\begin{equation*}
    \maximize_{\pi\in \Pi} \quad  \frac 1 \beta \Ep{x\sim \pi}{\exp(\beta R(x))}.
\end{equation*}
In addition to this connection,
exponential criteria are well-understood 
in the context of risk-sensitive control
\cite{james1994risk,baras1994robust,baras1998robust},
%
where the equivalence of the problem of robust output feedback control of general nonlinear, set valued, Markov chain, partially observed systems to 
a risk-sensitive partially observed control problem with an exponential of an integral cost criterion has been shown. 
In the case of known models, the solution has the known structure of two Hamilton-Jacoby-Bellman (HJB) equations: one forward in time that computes online the information state of the problem (i.e. the sufficient statistics for the control) and one backwards in time that computes off-line the control as a memory-less function of the information state. 
These results naturally suggest that, in the case of unknown models, reinforcement learning 
with exponential criteria may be introduced to potentially increase robustness in RL applications
using online data-driven estimation updates.

\subsection*{Contribution}

In this work, we study the effect of exponential criteria on the robustness of the learned policy of a reinforcement learning agent. In particular, we 

\begin{itemize}
    \item formulate the risk-sensitive reinforcement learning problem as an  
    optimization problem with a modified objective using exponential criteria and show its connection to KL-regularized RL methods (Section \ref{sSec:preliminaries-exp});
    \item provide a definition of robustness for RL policies and show that the use of exponential criteria results in robust RL policies with given probability bounds on the observed cumulative rewards in the form of concentration inequalities (Section \ref{sSec:preliminaries-robustness});
    \item provide new analytic results for the implementation of the risk-sensitive REINFORCE algorithm based on exponential criteria introduced in \cite{Noorani2021REINFORCE} regarding the update rule
    and the convergence of the parameters (Section \ref{Sec:R-PG});
    \item develop a risk-sensitive online actor-critic algorithm, by approximately solving a risk-sensitive multiplicative Bellman equation
    with stochastic approximation updates (Section \ref{Sec:AC}); 
    and
    \item quantify the robustness of the proposed methods in terms of the Conditional Value at Risk 
    (CVaR) values of the total reward
    in simulated experiments under model parameter perturbations
    (Section \ref{Sec:Experiments}).
\end{itemize}

Our experimental results support our theoretical analysis and suggest that the 
proposed problem formulation using exponential criteria is suitable for risk-sensitive 
reinforcement learning.
The proposed risk-sensitive RL methods inherit 
computational and convergence properties of widely-used RL algorithms, can
accelerate the learning process, and can reduce the variance of the learned policies under model uncertainty, 
resulting in policies that show enhanced robustness with respect to environmental and model perturbations.
\section{Risk-Sensitive Reinforcement Learning with Exponential Criteria}
\label{sec:preliminaries}

In this section, we formulate the problem of risk-sensitive reinforcement 
learning and show its connection to exponential criteria. 
We provide an explicit definition of robustness and risk-sensitivity,
and show that the use of exponential criteria is associated 
with a min-max optimization problem that results in robust RL policies
with known concentration bounds on the observed rewards under environmental perturbations.

\subsection{Reinforcement Learning Preliminaries}
\label{ssec:rl}

The RL problem is typically modeled using a Markov Decision Process (MDP) 
which is represented by a tuple 
$\mathcal{M} $$=$$ (\mathcal{S}, \mathcal{A}, p_0, P, r, \gamma)$, 
where $\mathcal{S}$ and $\mathcal{A}$ are, respectively, the state and action spaces (may each be discrete or continuous). 
The probability distribution $p_0$ is the initial state distribution, prescribing a probability on the starting state. The kernel $P : \mathcal{S} \times \mathcal{A} \rightarrow \Delta(\mathcal{S})$ is the transition kernel (unknown to the agent), where $\Delta(\mathcal{S})$ denotes the space of probability distributions on $\mathcal{S}$. 
The function $r : \mathcal{S} \times \mathcal{A} \rightarrow \mathbb{R}$ is the reward function; and $\gamma \in [0, 1)$ is a discounting factor. The behavior of an RL agent is determined by its policy. Here, we consider randomized policies. 

A (randomized) policy $\pi(\cdot|s)\in\Pi$ is a probability distribution over the action space given the state, which prescribes the probability of taking an action $a \in \mathcal{A}$ when in state $s$. Stochastic policies are smooth and continuous functions and therefore are more suitable for gradient-based methods.
At each time-step $t$, the agent perceives the state of the environment $s_t$, and executes an action $a_t$ according to its policy, a differentiable parametrized policy (e.g., a Neural Network), $\pi(\cdot|s_t; \theta)$ where $\theta \in \mathbb{R}^d$ is a vector of parameters. 
Then, the system transitions to a successor state $s_{t+1}$ according to transition probability $p(s_{t+1}|s_t,a_t)$ (unknown to the agent) and the agent receives a reward $r_t := r(s_t,a_t)$. 
By following policy $\pi$, the agent generates trajectory $\tau(\pi)$ (a sequence of states and actions). The agent's policy and the system transition probabilities induce a trajectory distribution, a probability distribution $\rho_\theta\in\Psi$ over the possible trajectories $\mathcal{T}$, given by
\begin{equation}
   \rho_{\theta}(\tau) = p_0 \prod_{t=0}^{|\tau|-1} \pi(a_t|s_{t}; \theta) p(s_{t+1}|s_t,a_t).
    \label{eq:rhotheta}
\end{equation}
%
%
The RL agent aims to find a policy that maximizes the sum of rewards over a time period, called episode. Since the observed rewards $r_t$ are random variables, 
in risk-neutral RL, the typical objective is to optimize for the
expected (discounted) cumulative reward:
\begin{equation} 
        \max_{\theta}. \  J(\theta) := \mathbb{E}_{\tau \sim \rho_{\theta}} \Big[R(\tau)\Big],\ 
        R(\tau) = \sum_{t=0}^{|\tau|-1} \gamma^t r(s_t, a_t), 
\label{eq:risk-neutral}
\end{equation} 
%
where $R(\tau)$ is the trajectory's ($\gamma$-discounted) total reward. 
The expectation is taken with respect to the trajectory distribution.  
That is,  the expectation is taken over the space of trajectories $\mathcal{T}$ generated by following the policy, 
i.e., $s_0 \sim p_0$, $a_t \sim \pi(\cdot | s_t; \theta)$ and $s_{t+1} \sim p(\cdot|s_t, a_t)$.
%

\subsection{Risk-Sensitive Reinforcement Learning}
\label{sSec:preliminaries-risk}

Risk-sensitivity in reinforcement learning is often associated 
with the following general problem:
\begin{equation} 
\max_{\theta} \quad  \inf_{\rho_{\theta}\in\Psi} \mathbb{E}_{\tau\sim\rho_\theta} \Big[R(\tau(\theta))\Big],
\label{eq:rsrl}
\end{equation} 
which induces distributional robustness with respect to the 
probability distribution over the possible trajectories $\mathcal{T}$.
Maximization over the parameter space $\theta \in \mathbb{R}^d$ simulates
optimization over all policies $\pi\in\Pi$.
Minimization over the distributions $\rho_\theta$ corresponds to reducing the sensitivity 
of the uncertainties that affect $\rho_\theta$, which include both 
the initial state distribution $p_0$,
and the transition probabilities $P$, i.e., all uncertainties with respect to the model parameters 
and any noise perturbation of the system dynamics.  
Typically the space $\Psi$ is constrained to a closed set of distributions that defines 
a trade-off between optimality and conservativeness of the policy.
However, solving \eqref{eq:rsrl} with dynamic programming and game theoretic methods becomes intractable in 
large state/action spaces, and methods that approximate its solution have been studied
\cite{tamar2013scaling,delage2010percentile,abdullah2019wasserstein},
including the use of different statistical measures of the objective function
to avoid the minimization over the distributions $\rho_\theta$ 
\cite{pan2019risk,dotan2012policy,prashanth2014policy,tamar2015risk,liu2018r2pg}. 

In this work, we focus on the following definition of a risk-sensitive reinforcement learning problem that incorporates an inherent regularization term
for the set of distributions $\rho_\theta$:
\begin{equation} 
\max_{\theta}. \quad 
\begin{cases}
    \sup_{\hat\rho} \cbra{ \Ep{\tau\sim \hat\rho}{R(\tau)} - \frac 1 \beta D_{KL}(\hat\rho,\rho_\theta) },\ \beta>0 \\
    \inf_{\hat\rho} \cbra{ \Ep{\tau\sim \hat\rho}{R(\tau)} - \frac 1 \beta D_{KL}(\hat\rho,\rho_\theta) },\ \beta<0
\end{cases}
\label{eq:rsrlkl}
\end{equation} 
where 
$D_{KL}(\cdot,\cdot)$ represents the Kullback-Leibler divergence measure defined in \eqref{eq:KL}.
%
%
The optimization problem \eqref{eq:rsrlkl} is essentially a game between 
the trajectory distribution $\hat\rho$ that tries to find a worst-case scenario of the cumulative reward while 
staying close to a baseline distribution
$\rho_\theta$, and the parameter vector $\theta$ that tries to optimize for 
the worst-case expected cumulative reward.
%
%
The use of baseline terms in reinforcement learning is widely adopted 
\cite{sutton2018reinforcement}
and is further explained in Section \ref{Sec:R-PG}. 
The parameter $\beta$ is the risk-sensitive parameter that controls the behavior of the agent and 
the weight of the regularization term. In particular, $\beta>0$ induces a risk-seeking (optimistic) approach, 
while $\beta<0$ invokes a risk-averse (pessimistic) approach
\cite{Noorani2021RR,Noorani2021risk}.
%
%

\subsection{Risk-Sensitive RL with Exponential Criteria}
\label{sSec:preliminaries-exp}

Problem \eqref{eq:rsrlkl} 
is a game-theoretic formulation of the risk-sensitive reinforcement learning problem, 
which can be hard to solve directly. 
However, it is well known (see, e.g., \cite{follmer2002convex},
), 
that the following duality relationship, with respect to a Legendre-type
transform, holds:  

\begin{theorem} 
Consider 
a measurable space
$(\Omega,\mathcal{F})$, where $\mathcal{F}$ is a $\sigma$-algebra on $\Omega$.
Let $\mathcal{P}(\Omega)$ be a set of probability measures $P:\Omega\rightarrow [0,1]$, 
and $P_\mu,P_\nu\in\mathcal{P}(\Omega)$.
In addition, consider a bounded measurable function $Z:\Omega\rightarrow\mathbb{R}$.
Then the free energy is defined as:
\begin{equation}
    J_{l_\beta} (Z) = \frac 1 \beta \log \Ep{P_\mu}{e^{\beta Z}} 
\end{equation}
and the KL divergence measure:
\begin{equation}
    D_{KL}(P_\nu,P_\mu) = \begin{cases}
    \int \log(\frac{dP_\nu}{dP_\mu})dP_\nu &\text{if } C_{KL}(P_\nu, P_\mu) \\
    \infty & \text{ otherwise }
    \end{cases} 
    \label{eq:KL}
\end{equation}
are in duality with respect to a Legendre-type transform, 
in the following sense:
\begin{equation}
    J_{l_\beta} (Z) = \begin{cases}
    \sup_{P_\nu\in\mathcal{P}(\Omega)} \cbra{ \Ep{P_\nu}{Z} - \frac 1 \beta D_{KL}(P_\nu,P_\mu) },\ \beta>0 \\
    \inf_{P_\nu\in\mathcal{P}(\Omega)} \cbra{ \Ep{P_\nu}{Z} - \frac 1 \beta D_{KL}(P_\nu,P_\mu) },\ \beta<0
    \end{cases}
    \label{eq:duality}
\end{equation}
Here the conditions $C_{KL}(P_\nu, P_\mu)$ include $P_\nu \ll P_\mu$ and 
$\int \log(\frac{dP_\nu}{dP_\mu})dP_\nu \in L^1(P_\nu)$.
\label{thm:duality}
\end{theorem}

\begin{proof}
    Follows directly from Theorem 6 in \cite{follmer2002convex} and 
    standard algebraic manipulations.
\end{proof}

\noindent
Corollary \ref{clr:rsrl} then follows directly from Theorem \ref{thm:duality}.

\begin{corollary} 
The problem:
\begin{equation} 
\max_{\theta}. \quad  J_{l_\beta}(\theta) := \frac 1 \beta \log \mathbb{E}_{\tau \sim \rho_\theta} \Big[\exp(\beta R(\theta))\Big]  
\label{eq:rsrllogexp}
\end{equation} 
is equivalent to \eqref{eq:rsrlkl}, for the baseline distribution $\rho_\theta$ being 
the current trajectory distribution of the algorithm, assuming that the maximum is attained.
\label{clr:rsrl}
\end{corollary}

Notice that a Taylor expansion of \eqref{eq:rsrllogexp} reveals an intuition behind 
how the exponential criterion incorporates risk into the objective function,
since it incorporates an infinite sum of the higher moments of the return, i.e.,
for small $\beta$ we get:
\begin{align}
    \frac 1 \beta \log \mathbb{E}\Big[ e^{\beta R(\theta)}\Big] = \mathbb{E}& \Big[R(\theta) \Big] + \frac{\beta}{2} \text{Var} \Big[R(\theta)\Big] + \mathcal{O}(\beta^2)
    \label{eq:exp-taylor}
\end{align}
Equation \eqref{eq:exp-taylor} shows how the risk-sensitive parameter $\beta$ 
controls the weight of the moments of the cumulative reward in the objective function. 
%
%
We note that as the risk-sensitive parameter $\beta$ approaches zero, the exponential objective \eqref{eq:exp-taylor} approaches the risk-neutral objective \eqref{eq:risk-neutral}. 
%

\begin{remark} Connection to Maximum-Entropy RL (see, e.g., \cite{eysenbach2021maximum}). 
Notice that we can simplify \eqref{eq:duality} by making heuristic assumptions on the measure $P_\mu$. 
In particular, it is known that the maximum-entropy objective is equivalent to the maximization of the KL-regularized objective 
with respect to a uniform distribution as the reference policy \cite{Galashov2019}.
In other words, by assuming that $P_\mu$ is a uniform probability measure, and for $\beta>0$, 
\eqref{eq:duality} and \eqref{eq:rsrllogexp} imply that
the problem of maximizing $J_{l_\beta} (\theta)$ with respect to $\theta$ is equivalent to 
\begin{equation*}
    \max_\rho.\ \cbra{ \Ep{\tau\sim\rho}{R(\tau)} + \frac{1}{\beta} H(\rho) },
\end{equation*}
where $H(\rho)$ represents the Shannon entropy of the distribution $\rho$.
Thus, the maximum-entropy RL objective of \cite{eysenbach2021maximum} is a special case of 
the objective \eqref{eq:rsrllogexp} considered in this work.
\end{remark}

\subsection{Policy Robustness}
\label{sSec:preliminaries-robustness}

Given an MDP $\mathcal{M} $$=$$ (\mathcal{S}, \mathcal{A}, p_0, P, r, \gamma)$
with transition probabilities $P$,
a fixed policy $\pi$, parameterized by $\theta$, defines
a trajectory distribution $\rho_\theta$ given 
by \eqref{eq:rhotheta}.
RL algorithms try to find the optimal policy $\pi(\theta)$ given observations of the rewards $r$ of $\mathcal M$.
However, during the testing phase when the policy $\pi(\theta)$ is applied, 
environment and model perturbations may alter the transition probabilities.
Thus, the agent is asked to operate on a perturbed MDP 
$\mathcal{\hat M} $$=$$ (\mathcal{S}, \mathcal{A}, \hat p_0, \hat P, r, \gamma)$,
where $\hat P$ represents the perturbed system of transition probabilities.
This is especially the case when training takes place in simulation environments
while testing is transferred to a real system.

In this case, risk-sensitivity can be associated with a measure of robustness of a policy $\pi(\theta)$,
quantified by a lower bound on the probability of good performance when the
transition distribution $\hat\rho$ during testing 
deviates from the distribution $\rho_\theta$
induced by $\pi(\theta)$.
In this work, we will adopt the following definition of robustness of a policy $\pi(\theta)$.

\begin{definition}
Let $\pi(\theta)$ be a given policy and $\rho_\theta$ be its associated trajectory distribution given 
by \eqref{eq:rhotheta} with transition probabilities $P$.
In addition, let $\hat\rho$ be a trajectory distribution  generated by $\pi(\theta)$
given a perturbed system of transition probabilities 
$\hat P$.
The policy $\pi(\theta)$ is $(\xi,\delta,\epsilon)$-robust if, 
for $\delta,\epsilon >0$, and
under the condition $D(\hat\rho,\rho_\theta) \leq \epsilon$, it holds that
\begin{equation}
    \mathbb{P}_{\tau \sim \hat\rho} \sbra{ R(\tau(\theta)) > \xi } \geq 1 - \delta(\xi,\epsilon), 
    \label{eq:gammaepsilondelta}
\end{equation}
where $D(\cdot,\cdot)$ represents the KL divergence defined in \eqref{eq:KL}.
\label{def:robust}
\end{definition}

In general, non-trivial sets of parameters $(\xi, \epsilon, \delta)$ such that the condition \eqref{eq:gammaepsilondelta} is satisfied cannot be found.
However, for optimal policies with respect to \eqref{eq:rsrllogexp}, 
we can analytically provide such parameters using standard concentration inequalities.
Theorem \ref{thm:upper-bounds} provides upper bounds on the probability of the tails
of the cumulative rewards $R$, in the case of bounded reward ($R\leq R_{max}$
almost surely). Note that $R_{max}= \frac{r_{max}(1-\gamma^T)}{1-\gamma}$ when the per step reward is bounded $r \leq r_{max}$.

\begin{theorem} 
Let $\pi(\theta^*)$ be an optimal policy
with respect to \eqref{eq:rsrllogexp}, i.e., $\pi(\theta^*)=\argmax_\theta J_{l_\beta}(\theta)$, and $\rho_{\theta^*}$ be its associated trajectory distribution  given by \eqref{eq:rhotheta} with transition probabilities $P$.
In addition, let $\hat\rho$ be a trajectory distribution generated by $\pi(\theta)$
given a perturbed system of transition probabilities $\hat P$ such that 
$D(\hat\rho,\rho_{\theta^*}) \leq \epsilon$.
Then the following inequalities hold:
\begin{align}
    &\mathbb{P}_{\tau \sim \hat\rho} \sbra{R(\tau) \geq \xi} 
    \leq \frac{1}{\xi}J_{l_\beta}^* + \frac{\epsilon}{\beta \xi},\ \beta>0,
    \label{eq:upper-bound-seeking} \\
    &\begin{aligned}
    &\mathbb{P}_{\tau \sim \hat\rho} \sbra{R(\tau) \leq \xi} \leq \\
    &\qquad \leq \frac{R_{max}}{R_{max}-\xi}\pbra{1 - \frac{1}{R_{max}}J_{l_\beta}^* + \frac{\epsilon}{|\beta| R_{max}}},\ \beta<0, 
    \end{aligned}
    \label{eq:upper-bound-averse} 
\end{align}
where $J_{l_\beta}^* = \frac 1 \beta \log \mathbb{E}_{\tau \sim \rho_{\theta^*}} \Big[\exp(\beta R(\tau))\Big]$. 
\label{thm:upper-bounds}
\end{theorem}
\begin{proof}
For \eqref{eq:upper-bound-seeking}, using Markov's inequality, we get:
\begin{equation}
\begin{aligned}
\mathbb{P}_{\tau \sim \hat\rho} \sbra{ R(\tau) \geq \xi } 
    &\leq \frac{\mathbbm E_{\tau \sim \hat\rho}[R(\tau)]}{\xi} \\
    &\leq \frac{1}{ \xi} \pbra{ J_{l_\beta}^* + \frac{1}{\beta} D(\hat\rho,\rho_{\theta^*}) } \\
    &\leq \frac{1}{\xi}J_{l_\beta}^* + \frac{\epsilon}{\beta \xi},  
\end{aligned}    
\end{equation}
where we have used \eqref{eq:duality} 
for $\beta>0$ to get: 
\begin{align*}
J_{l_\beta}^* &= \frac 1 \beta \log \mathbb{E}_{\tau \sim \rho_{\theta^*}} \Big[\exp(\beta R(\tau))\Big] \\ 
&= \sup_\rho \cbra{ \Ep{\tau\sim\rho}{R(\tau)} - \frac 1 \beta D(\rho,\rho_{\theta^*}) } \\
&\geq \Ep{\tau\sim\hat\rho}{R(\tau)} - \frac 1 \beta D(\hat\rho,\rho_{\theta^*}),
\end{align*}
which implies that $\Ep{\tau\sim\hat\rho}{R(\tau)} \leq  J_{l_\beta}^* + \frac 1 \beta D(\hat\rho,\rho_{\theta^*})$.

Similarly, for \eqref{eq:upper-bound-averse}, using reverse Markov's inequality and 
assuming that $R<R_{max},\ a.s.$, we get:
\begin{equation}
\begin{aligned}
\mathbb{P}_{\tau \sim \hat\rho} &\sbra{ R(\tau) \leq \xi } 
    \leq \frac{R_{max} - \mathbbm E_{\tau \sim \hat\rho}[R(\tau)]}{R_{max}-\xi} \\
    &\leq \frac{R_{max}}{R_{max}-\xi} \pbra{ 1 - \frac{1}{R_{max}} J_{l_\beta}^* - \frac{1}{R_{max} \beta} D(\hat\rho,\rho_{\theta^*}) } \\
    &\leq \frac{R_{max}}{R_{max}-\xi}(1 - \frac{1}{R_{max}}J_{l_\beta}^* + \frac{\epsilon}{|\beta| R_{max}})  
\end{aligned}    
\end{equation}
where we have used \eqref{eq:duality} 
for $\beta<0$ to get: 
\begin{align*}
J_{l_\beta}^* &= \frac 1 \beta \log \mathbb{E}_{\tau \sim \rho_{\theta^*}} \Big[\exp(\beta R(\tau))\Big] \\
&= \inf_\rho \cbra{ \Ep{\tau\sim\rho}{R(\tau)} - \frac 1 \beta D(\rho,\rho_{\theta^*}) } \\
&\leq \Ep{\tau\sim\hat\rho}{R(\tau)} - \frac 1 \beta D(\hat\rho,\rho_{\theta^*}),\ \beta<0,
\end{align*}
which implies that $-\Ep{\tau\sim\hat\rho}{R(\tau)} \leq  -J_{l_\beta}^* - \frac 1 \beta D(\hat\rho,\rho_{\theta^*})$.
\end{proof}

\begin{remark}
Note that in Theorem \ref{thm:upper-bounds}, the term $J_{l_\beta}^*$
does not depend on the perturbed system of transition probabilities $\hat P$ in the test environment.
\end{remark}

Equations \eqref{eq:upper-bound-seeking} and \eqref{eq:upper-bound-averse} give upper bounds
on the probability of the two tails of the cumulative rewards $R$.
In particular, a risk-averse agent tries to optimize for 
the maximum average reward weighing in the maximization of the decay of  
the left tail of the distribution of the total reward, while 
a risk-seeking agent weights in the maximization of the decay of the right tail
of the reward distribution. 
This is consistent with the following theorem proven in 
\cite{Noorani2021RR} using the Gartner-Ellis theorem of Large Deviation: 

\begin{theorem} [\cite{Noorani2021RR}]
For a given negative risk parameter (risk-aversion) $\beta $$<$$ 0$, the maximization of the risk-sensitive exponential criterion $J_{l_\beta}$ in \eqref{eq:rsrllogexp} is equivalent to the maximization of the exponential rate of decay of the left tail of the system's trajectory reward distribution, i.e., for a given $\beta<0$, there exists a constant $\psi $$\in$$ \mathbb{R}$ such that
%
%
\begin{align*}
    \argmax_{\pi} J_{l_\beta}(\pi) = \lim_{|\tau| \to \infty} \argmin_{\pi} \mathbb{P}[R(\tau) < \psi],
\end{align*}
where $\mathbb{P}[R(\tau) < \psi]$ denotes the probability of the event $R < \psi$. 
Similarly, for a given positive risk parameter (risk-seeking) $\beta $$>$$ 0$, the maximization of the risk-sensitive exponential criterion $J_{l_\beta}$ in \eqref{eq:rsrllogexp} is equivalent to minimization of the exponential rate of decay of the right tail of the system's trajectory reward distribution, that is, for a given $\beta>0$, there exists a constant $\psi$ such that
\begin{align*}
    \argmax_{\pi} J_{l_\beta}(\pi) = \lim_{|\tau| \to \infty} \argmax_{\pi} \mathbb{P}[R(\tau)) > \psi].
\end{align*}
\label{thm:ldt}
\end{theorem}

\noindent
Based on Theorem \ref{thm:upper-bounds}, 
Corollary \ref{cor:robust-averse} shows that the risk-averse 
policy ($\beta<0$) with respect to \eqref{eq:rsrllogexp} is a $(\xi,\delta,\epsilon)$-robust policy according to 
Definition \ref{def:robust}.

\begin{corollary}
Let an optimal policy $\pi(\theta^*)=\argmax_\theta J_{l_\beta}(\theta)$ 
with respect to \eqref{eq:rsrllogexp} for $\beta<0$.
%
%
Then, $\pi(\theta^*)$ is $(\xi,\delta,\epsilon)$-robust according to 
Definition \ref{def:robust} with:
\begin{align}
    \delta(\xi,\epsilon) = \frac{R_{max}}{R_{max}-\xi}(1 - \frac{1}{R_{max}}J_{l_\beta}^* + \frac{\epsilon}{|\beta| R_{max}}).
    \label{eq:delta} 
\end{align}
In addition, for a given $\delta = \bar{\delta}$, we can quantify $\xi$ by:
    \begin{align}
         \xi = R_{max}  - \frac{R_{max}}{\bar{\delta}} (1-J_{\beta}^*+\frac{\epsilon}{\beta R_{max}}).
    \end{align}
\label{cor:robust-averse}
\end{corollary}
\begin{proof}
    It follows from \eqref{eq:upper-bound-averse} since 
    $\mathbb{P}[R > \xi] $$=$$ 1 - \mathbb{P}[R\leq \xi]$.
\end{proof}

As a last remark, we have shown how the optimization problem \eqref{eq:rsrllogexp}
is connected to risk-sensitivity and robustness of the learned policy with respect to 
model perturbations.
However, as will be discussed in Section \ref{sSec:RSactorcritic}, 
the presence of the logarithmic non-linearity 
in \eqref{eq:rsrllogexp} creates computational problems in 
algorithmic implementations.
For this reason, throughout the rest of this paper 
we will study the equivalent (in terms of optimal policy) 
problem:
\begin{equation} 
\max_{\theta}.\ J_\beta(\theta) := \frac 1 \beta \mathbb{E}_{\tau \sim \rho_\theta} \Big[ \exp(\beta R(\theta))\Big].
\label{eq:rsrlexp}
\end{equation} 
%



\section{Policy Gradient with Exponential Criteria}
\label{Sec:R-PG}

In this section, we present a brief overview of the 
risk-sensitive REINFORCE algorithm introduced in \cite{Noorani2021REINFORCE} and provide new analytic results for its implementation 
regarding its update rule and the convergence of its parameters.
%

\subsection{Policy Gradient and the REINFORCE Method}
\label{sSec:REINFORCE}

Policy Gradient (PG) methods are a class of Policy Search methods that use gradient ascend/descend schemes to search for the optimal policy \cite{sutton2000policy,kakade2001natural,gu2017interpolated}.
That is, at each iteration of the algorithm $t$, the parameters of the policy are updated using the following update rule
\begin{align} \label{Eq:PG}
    \theta_{t+1} = \theta_{t} + \alpha \widehat{\nabla J(\theta_t)}
\end{align}
where $\alpha $$\in$$ \mathbb{R}$ is a constant step-size, i.e., learning rate, and $\widehat{\nabla J(\theta_t)} $$\in$$ \mathbb{R}^d$ is an unbiased estimate of the gradient with respect to the policy parameter $\theta$.
The well-known REINFORCE \cite{williams1991} and Actor-Critic \cite{konda1999actor} algorithms are examples of monte-carlo and recursive on-policy policy gradient algorithms, respectively, particularly suitable for continuous action spaces.

An estimate of the gradient of $J$ in \eqref{Eq:PG} with respect to the policy parameters can be obtained using the policy gradient theorem \cite{sutton1999policy}, that is,
\begin{equation} \label{grad:risk-neutral_vanila}
    \begin{aligned}
    \nabla J (\theta) \propto \mathbb{E}_{\tau \sim \rho_{\theta}}\Big[ R(\tau) \sum_{t=0}^{|\tau|-1}\nabla\log \pi_{\theta}(a_t|s_t; \theta)\Big].
    \end{aligned}
\end{equation}
The policy gradient theorem suggests that the gradient estimate in \eqref{Eq:PG} can be computed by Monte Carlo estimation of the expectation in \eqref{grad:risk-neutral_vanila}. 
%
Eq. \eqref{grad:risk-neutral_vanila} can be re-written in terms of the reward-to-go $R_t := \sum_{t'=t}^{|\tau|-1} \gamma^{t'-t} r(s_{t'},a_{t'})$ as follows \cite{sutton2018reinforcement}:
\begin{align} \label{grad:risk-neutral_rtg}
    \nabla J (\theta) \propto \mathbb{E}_{\tau \sim \rho_{\theta}}\Big[ \sum_{t=0}^{|\tau|-1} R_t \nabla\log \pi_{\theta}(a_t|s_t; \theta)\Big].
\end{align}
Using Eq. \eqref{grad:risk-neutral_rtg}, the update rule in the standard REINFORCE algorithm is obtained and is given by
\begin{equation}
    \begin{aligned}
    \theta_{t+1} = \theta_t + \alpha R_t \frac{\nabla \pi(a_t|s_t;\theta)}{\pi(a_t|s_t;\theta)}.
    \end{aligned}
\end{equation}
%
%
%
%
%
%
To further reduce the variance associated with the gradient estimations of \eqref{grad:risk-neutral_vanila} and \eqref{grad:risk-neutral_rtg}, which is imperative in complex environments, baseline methods, based on subtracting an appropriately chosen baseline from the reward-to-go $R_t$, have been proposed.
Using baselines, one gets
\begin{equation}
    \begin{aligned}
    \nabla J (\theta) \propto \mathbb{E}_{\tau \sim \rho_{\theta}}\Big[ \sum_{t=0}^{|\tau|-1} \Big(R_t -b(s_t)\Big) \nabla\log \pi_{\theta}(a_t|s_t; \theta)\Big]
    \end{aligned}
\end{equation}
where $b(s_t)$ is a state-dependent 
function \cite{sutton2018reinforcement}. State-dependent baselines are guaranteed to exist, introduce no bias, and 
show better convergence properties in practice.
However, they are hard to find \cite{weaver2013optimal}. A common baseline in practice is the estimate of the value function, i.e., $b(s_t) = V^{\pi_{\theta}}(s_t):= \mathbb{E}_{\tau \sim \rho_{\theta}}\Big[R_t|s_t\Big]$.
%
%
As we will show, 
a particularly convenient property of using 
exponential criteria is that it alleviates the need for such approaches \cite{Noorani2021risk}.
%

%

\subsection{Risk-sensitive REINFORCE (R-REINFORCE)}
\label{sSec:RSreinforce}

Risk-sensitive REINFORCE (R-REINFORCE) \cite{Noorani2021REINFORCE} 
is a the risk-sensitive counterpart of REINFORCE based on the objective \eqref{eq:rsrlexp}.
In R-REINFORCE, the update rule \eqref{grad:risk-neutral_rtg} is replaced by:
\begin{equation} 
    \begin{aligned}
        \nabla J_{\theta}(\theta) \propto \frac 1  \beta \mathbb{E}_{\tau \sim \rho_{\theta}} \Big[ \sum_{t=0}^{|\tau|-1} e^{\beta R_t} \nabla \log \pi_t(\theta)\Big].
    \end{aligned}
    \label{eq:rspgt}
\end{equation}
%
The derivation of this formula is based on a risk-sensitive variation of the 
policy gradient theorem \cite{sutton1999policy}.
These results are provided
in Appendix \ref{apndx:derivation}. 
Given \eqref{eq:rspgt}, the R-REINFORCE update rule reads as: 
\begin{equation} \label{update_rule_exp}
    \begin{aligned}
    \theta_{t+1} = \theta_t + \frac \alpha \beta e^{\beta R_t} \frac{\nabla \pi(a_t|s_t; \theta)}{\pi(a_t|s_t; \theta)}.
    \end{aligned}
\end{equation}
and is a stochastic approximation algorithm 
(see, e.g., \cite{borkar2009stochastic}).
We provide the convergence analysis of the parameters $\theta$ 
in Appendix \ref{apndx:convergence}.
The implementation of the Risk-sensitive REINFORCE algorithm is given in Alg. \ref{alg:RiskSensitiveREINFORCE}.
%
For more details, the readers are referred to \cite{Noorani2021REINFORCE}
and the references therein.

\begin{algorithm}[h]
  \caption{Risk-sensitive REINFORCE}
  \label{alg:RiskSensitiveREINFORCE}
\begin{algorithmic}[1]
  \STATE {\bfseries Input:} a differentiable policy $\pi(a|s;\theta)$.
  \STATE {\bfseries Algorithm parameters:} step-size $\alpha$$>$$0$, \\discount factor $\gamma$$>$$0$, risk parameter $\beta$.
  \STATE {\bfseries Initialization:} $\theta=\theta_0 \in \mathbb{R}^d$.
  \WHILE {\bfseries $\theta$ not converged} 
  \STATE Generate an episode $s_0,a_0,$$\dots$$,s_{|\tau|-1}, a_{|\tau|-1}$  \\
  by $s_0 $$\sim$$ p_0$, $a_t $$\sim$$ \pi(\cdot | s_t; \theta)$, $s_{t+1} $$\sim$$ p(\cdot|s_t, a_t)$
  \FOR{$t=0$ {\bfseries to} $|\tau|-1$}
  \STATE $\hat R \leftarrow \sum_{t'=t}^{|\tau|-1} \gamma^{t'-t} r_t$
  \STATE $\theta_{t+1} \leftarrow \theta_t + \alpha \gamma^t \ \frac 1 \beta  e^{\beta \hat R} \ \nabla\log \pi(a_t|s_t; \theta_t)$
  \ENDFOR
  \ENDWHILE
\end{algorithmic}
\end{algorithm}

\noindent
Note that the update rule is not proportional to the reward-to-go $R_t := \sum_{t'=t}^{|\tau|-1} \gamma^{t'-t} r(s_{t'},a_{t'})$, but to the exponential%
\begin{equation}
\beta e^{\beta R_t} = \frac 1 \beta \prod_{t'=t}^{|\tau|-1} \exp\{\gamma^{t'-t} \beta r(s_{t'},a_{t'})\}. 
\label{eq:product}
\end{equation}
%

\begin{remark}
Substituting the exponential with its Taylor series expansion (see eq. \eqref{eq:exp-taylor}), 
reveals that the risk-sensitive objective provides a natural baseline (see Section \ref{sSec:REINFORCE}). 
This is empirically shown in \cite{Noorani2021REINFORCE}. The baseline term can be derived by expanding the exponential function and combining all terms, except for the one proportional to $R_t$, i.e., $\nabla J (\theta) \propto \mathbb{E}_{\tau \sim \rho_{\theta}}\Big[ \sum_{t=0}^{|\tau|-1} \Big(R_t -b(s_t)\Big) \nabla\log \pi_{\theta}(a_t|s_t; \theta)\Big]$ where $b(s_t) = -(\frac{1}{\beta} + \frac{\beta R_t^2}{2}+\dots)$.
%
In Section \ref{Sec:Experiments}, we show that such baseline leads to significant variance reduction and acceleration of learning process.
\end{remark}

\section{Actor-Critic with Exponential Criteria}
\label{Sec:AC}

Actor-Critic methods \cite{konda1999actor,konda2003actor,la2013actor,bhatnagars2009} improve the policy using gradient methods and use a critic network to estimate the value function and use it to bootstrap an estimate of the reward-to-go. 
%
%
The value function
$V^{\pi_\theta}(s_t)\simeq \mathbb{E}_{\tau \sim \rho_{\theta}}[R_t|s_t]$,
satisfies the Bellman's equation
\begin{equation}
\begin{aligned}
V^{\pi_{\theta^*}}(s)=  \mathbb{E}_{a \sim \pi_{\theta^*}}\Big[ r(s,a) 
+ \gamma V^{\pi_{\theta^*}}(s') |s \Big], 
\end{aligned}
\label{eq:HJB}
\end{equation}
%
%
which is a contraction mapping that gives rise to 
stochastic approximation algorithms that 
try to asymptotically minimize the mean-squared error
\begin{align*}
\minimize_{\theta} ~ 
  \mathbb{E}_{a \sim \pi_{\theta}}\Big[ \| r(s,a) + 
    \gamma V^{\pi_{\theta}}(s') - V^{\pi_{\theta}}(s) \|^2 \mid s \Big],   
\end{align*}
forming temporal-difference actor-critic methods that employ learning models 
(e.g. neural networks \cite{gu2016continuous} or other models \cite{mavridis2021vector,mavridis2021maximum}).
%
%
%
Such methods
use two learning systems to estimate the parameters $\theta_t$
of the optimal policy $\pi(a_t|s_t; \theta_t)$ (actor) and 
the parameters $w_t$ of the value function $V(s_t;w_t)$ (critic), that is
%
%
\begin{equation}
\begin{cases}
    \theta_{t+1} = \theta_{t}+\alpha\Big(\hat R_t -V(s_t;w_t)\Big) 
    \frac{\nabla \pi(a_t|s_t;\theta_t)}{\pi(a_t|s_t; \theta_t)}  \\
    w_{t+1} = w_{t} - \bar{\alpha} \nabla J_c(s_t;w_t,\theta_t)
\end{cases},
\label{eq:ac}
\end{equation}
where $J_c(s_t;w_t,\theta_t)= \| \hat R_t - V(s;w_t) \|^2$.
In this case, $\hat R_t$ is an estimate of the reward-to-go $R_t$ given by  
\begin{equation*}
    \hat R_t:= r(s,\pi_{\theta_t}) + \gamma V(s';w_t) 
\end{equation*}
\subsection{Risk-Sensitive Online Actor-Critic (R-AC)}
\label{sSec:RSactorcritic}

In this section, we develop a risk-sensitive counterpart of the temporal-difference actor-critic method.
%
%
%
In contrast to the risk-neutral case, 
in the risk-sensitive reinforcement learning setting 
the optimal control problem is often 
associated with an undiscounted version of the
cost function $J_{l_\beta}$ in \eqref{eq:rsrllogexp}:
\begin{equation} 
\maximize_{\pi} \bar J_{l_\beta}(\pi) := \limsup_{n\rightarrow\infty} \frac 1 n \log \E{ e^{ \beta \sum_{l=0}^{n-1} r(s_l,a_l)} | s_0}.
\label{eq:rsHJBlog}
\end{equation} 
Notice that it has been assumed that $\gamma=1$,
and the time-average limit has been added to ensure boundedness of the cost.
It has been shown (see, e.g., \cite{borkar2002q,borkar2010learning})
that by defining a value function  
$\bar V_{l_\beta}^*(s_k)= \max_{\pi} \E{ e^{ \beta \sum_{l=k}^{t_h} r(s_l,a_l) - \log J_{l_\beta}^* }|s_k } $
,  
with $t_h$ being the first hitting time of a distinguished state,
problem \eqref{eq:rsHJBlog} is equivalent to a multiplicative version of 
the Bellman equation
which defines a nonlinear eigenvalue problem:
\begin{equation} 
\bar V_{l_\beta}^*(s_k) = \max_{\pi} \frac{e^{ \beta r(s_k,a_k)} }{\bar J_{l_\beta}^*} \E{ V_{l_\beta}^*(s_{k+1}) | s_k}
, \quad a_k \sim \pi(\cdot | s_k).
\label{eq:rsHJBlogeig}
\end{equation} 
For sufficiently small $\beta$, 
stochastic approximation updates in two timescales can be designed to 
solve the eigenvalue problem recursively implementing a policy iteration scheme
and converging to 
an optimal stationary control that attains the optimal reward $J_{l_\beta}^*<\infty$. 
It is important to point out that 
substituting for the logarithmic value function $W(\cdot) = \log V_{l_\beta}(\cdot)$
results in an additive dynamic programming equation,
that has similarities with the classical 
equation for average reward:
{\small
\begin{equation}
W^*(s_k):=\max_{\pi} \cbra{r(s_k,a_k) + \log \E{e^{W^*(s_{k+1})} \mid s_k}} -\log J_{l_\beta}^*.
\label{eq:notHJB}
\end{equation}
}%
%
While this seems like a compelling formulation, and has indeed been followed by some authors
(see, e.g., \cite{fei2020risk,fei2021exponential}), 
the problem arises when attempting 
to formulate a reinforcement learning algorithm out of the latter dynamic programming equation. 
In particular, notice that, in eq. \eqref{eq:notHJB}, the conditional expectation 
with respect to the transition probabilities appears inside a logarithm, 
in contrast to eq. \eqref{eq:multiBellman}.
This typically leads to violation of the assumptions of the stochastic approximation algorithm used to train 
temporal-difference RL algorithms (e.g., stochastic gradient descent if using neural networks) \cite{borkar2010learning}.
As a result, the form of eq. \eqref{eq:notHJB} is not convenient for Q-learning
and most temporal-difference RL methods.

In this work, we consider the discounted optimal control problem in
%
%
\eqref{eq:rsrlexp}.
According to the cost function $J_\beta$, 
we define the risk-sensitive value function of a policy $\pi$ as
$V_\beta^\pi(s_k) := \frac 1 \beta \E{ e^{ \beta \sum_{l=k}^\infty \gamma^{l-k} r(s_l,a_l)} |s_k }$.
%
%
%
We further define: 
\begin{equation}
\bar V_\beta^\pi(s_k) := \beta V_\beta^\pi(s_k) = \E{ e^{ \beta \sum_{l=k}^\infty \gamma^{l-k} r(s_l,a_l)} |s_k }.
\end{equation}
By definition, we get that $\bar V_\beta^\pi(\cdot)\geq 0$, and the following 
relationship holds:
%
\begin{align}
\bar V_\beta^*(s_k) :&= \max_{\pi}~ \E{ e^{ \beta \sum_{l=k}^\infty \gamma^{l-k} r(s_l,a_l)}|s_k } \nonumber
	\\&= \max_{\pi}~  \E{ e^{ \beta \pbra{ r(s_k,a_k) + \gamma \sum_{l=k+1}^\infty \gamma^{l-(k+1)} r(s_l,a_l)} } |s_{k}} \nonumber 
 \\&= \max_{\pi}~ e^{\beta r(s_k,a_k)} \E{ (\bar V_\beta^{*})^\gamma (s_{k+1})|s_{k} }  + \bar\epsilon_k(\gamma) \nonumber 
 \\&= \max_{\pi}~ \E{ e^{\beta r(s_k,a_k) + \gamma \log \bar V_\beta^{*} (s_{k+1})}|s_{k} } + \bar\epsilon_k(\gamma)
\label{eq:multiBellman}
\end{align}
where $\bar V^*(\cdot) = \bar V^{\pi^*}(\cdot)$ is the optimal value function resulting by the optimal control policy, and the term $\bar\epsilon(\gamma)$ is given by:
\begin{equation}
\begin{aligned}
    \bar\epsilon_k(\gamma) &= e^{\beta r(s_k,a_k)} \mathbb E [ \pbra{e^{ \beta \sum_{l=k+1}^\infty \gamma^{l-(k+1)} r(s_l,a_l)}}^\gamma 
    \\&\qquad\qquad\quad - \E{e^{ \beta \sum_{l=k+1}^\infty \gamma^{l-(k+1)} r(s_l,a_l)}|s_{k+1}}^\gamma  |s_{k} ]
    \\&= e^{\beta r(s_k,a_k)} \mathbb E[ e^{ \gamma \sum_{l=k+1}^\infty \gamma^{l-(k+1)} r(s_l,a_l)} \\&\qquad\qquad\qquad\qquad\qquad\qquad\qquad\ - (\bar V_\beta^{*})^\gamma (s_{k+1})|s_{k} ]
\end{aligned}
    \label{eq:multiBellmanError}
\end{equation}
Note that the existence of the term $\bar\epsilon(\gamma)$ implies
that \eqref{eq:multiBellman} holds only approximately. 
The approximation error $\bar\epsilon(\gamma)$ depends on the statistics of the problem at hand, 
as well as the value of $\gamma$. 
A good approximation can be achieved for $\gamma\approx 1$, since
strict equality in \eqref{eq:multiBellman} holds in the case 
of $\gamma=1$, when $\bar\epsilon_k(\gamma) = 0$, $\forall k\geq 0$.
This follows from the law of total expectation such that  
$\E{ e^{ \beta \sum_{l=k+1}^\infty r(s_l,a_l)} |s_{k}}$ $=$
$\E{\E{e^{ \beta \sum_{l=k+1}^\infty r(s_l,a_l)} |s_{k+1}}|s_{k}}$ $=$ 
$\E{ (\bar V_\beta^{*}) (s_{k+1})|s_{k} }$.
Notice also how the use of the exponential has resulted in a multiplicative Bellman equation.
%
Finally, note that the exponent $\gamma$ 
is assumed a rational number such that the term $(\bar V_\beta^{*})^\gamma$ is well-defined.
%
%
This is not restrictive, as in practice the term $\exp (\gamma \log \bar V_\beta^{*})$ is used, 
leading to a similar update law to the risk-neutral case.
%



To develop a risk-sensitive temporal-difference reinforcement learning algorithm, 
we use two learning systems, similar to \eqref{eq:ac}, as follows:
\begin{equation}
\begin{cases}
    \theta_{t+1} = \theta_{t}+\alpha \frac{1}{|\beta|}\big( R_t^\beta - \bar V_\beta (s_t;w_t) \big) 
    \frac{\nabla \pi(a_t|s_t;\theta_t)}{\pi(a_t|s_t; \theta_t)}  \\
    w_{t+1} = w_{t} - \bar{\alpha} \nabla J_r(s_t;w_t,\theta_t)
\end{cases}
\label{eq:rsacupdates}
\end{equation}
where, 
in contrast to the risk-neutral case in \eqref{eq:ac}, here we define
\begin{equation}
  R_t^\beta = \exp\sbra{ \beta r(s_t,a_t) + \gamma \log \bar V_\beta (s_{t+1};w_t) },
\end{equation}
\begin{equation}    
\begin{aligned}
J_r(s_t;w_t,\theta_t) 
      &= \| \exp\sbra{ \beta r(s_t,a_t) + \gamma \log \bar V_\beta (s_{t+1};w_t) } 
     \\ & \qquad - \bar V_\beta(s_t;w_t) \|^2
     ,\quad a\sim \pi_{\theta_{t}},
\end{aligned}
\label{eq:rsac}
\end{equation}
forming a stochastic gradient descent approach to asymptotically minimize 
the mean-squared error:
\begin{align*}
\min_{w} ~ 
  \E{ \| e^{\beta r(s_t,a_t)} (\bar V_\beta)^\gamma(s_{t+1};w)
     - \bar V_\beta(s_t;w) \|^2  \mid s_t}
\end{align*}
The actor parameter updates constitute a stochastic approximation algorithm 
based on \eqref{update_rule_exp}, where the average reward-to-go $V_\beta(s_t;w_t)=\frac 1 \beta \E{ e^{ \beta R_k} |s_k }$ is estimated
by the critic model. 
%
The critic parameter updates are also a stochastic approximation scheme 
that run in a slower timescale (see, e.g., \cite{borkar2010learning}).
Notice that this recursion does not correspond to a fixed-point iteration but to 
a stochastic gradient descent approach.
%
The algorithmic implementation is based on the updates \eqref{eq:rsacupdates} 
and the objective function in \eqref{eq:rsac} and is provided in Alg. \ref{alg:RiskSensitiveActorCritic}.
%
%


\begin{remark}
Note that simply minimizing the error
$\| \beta e^{ \beta r(s,a)} + \gamma V(s';w_t) - V(s;w_t) \|$, $a\sim \pi_{\theta_{t}}$,
for the risk-neutral value function $V$ 
is not equivalent to the update rule \eqref{eq:rsacupdates}, but to simply scaling the initial 
rewards $r_t$ to $\beta e^{\beta r_t}$.
%
\end{remark}


%
\begin{algorithm}[h]
  \caption{Risk-sensitive Online Actor-Critic (R-AC)}
  \label{alg:RiskSensitiveActorCritic}
\begin{algorithmic}[1]
  \STATE {\bfseries Input:} a differentiable policy $\pi(a|s;\theta)$.
  \STATE {\bfseries Algorithm parameters:}\\step-sizes $\alpha$$>$$0$, $\bar{\alpha}$$>$$0$, \\
  discount factor $\gamma$$>$$0$, risk parameter $\beta$.
  \STATE {\bfseries Initialization:} $\theta=\theta_0 \in \mathbb{R}^d$, $w=w_0 \in \mathbb{R}^{d'}$.
  \WHILE {\bfseries $(\theta,w)$ not converged}
  \FOR{$t=0$ {\bfseries to} $|\tau|-1$}
  \STATE $a_t $$\sim$$ \pi(\cdot | s_t; \theta)$, $s_{t+1} $$\sim$$ p(\cdot|s_t, a_t)$ 
  \STATE $\hat R_\beta \leftarrow \beta r_t + \gamma \log \bar V_\beta(s_{t+1};w_t)$
  \STATE {\small $\theta_{t+1} \leftarrow \theta_t + {\alpha \gamma^t \frac{1}{|\beta|} (e^{\hat R_\beta} - \bar V_\beta(s_t;w_t)) \nabla\log \pi(a_t|s_t; \theta)}$}
  \STATE $w_{t+1} \leftarrow w_{t} + \bar{\alpha} \gamma^t (e^{\hat R_\beta} - \bar V_\beta(s_t;w_t)) \nabla \bar V_\beta(s_t;w_t)$
  \ENDFOR
  \ENDWHILE
\end{algorithmic}
\end{algorithm}
%


\section{Simulation Results}
\label{Sec:Experiments}

To evaluate the proposed risk-sensitive reinforcement learning algorithms, we compare them against their risk-neutral counterparts on two classic reinforcement learning problems, namely the inverted pendulum (Cart-Pole) 
and the underactuated double pendulum (Acrobot) \cite{sutton2018reinforcement}
. 
The experiments are designed to investigate the performance and 
robustness of the proposed risk-sensitive algorithms against model perturbations.
We quantify the performance of the algorithms 
using the mean values of the 
the observed cumulative rewards $R$ during testing in different environments, 
and their robustness using the variance and the Conditional Value at Risk (CVaR)%
\footnote{Equation \eqref{eq:cvar} captures the intuition behind the statistical meaning of CVaR and
holds if there is no probability atom at
$VaR_p(R)$. For a formal definition the readers are referred to \cite{chow2014algorithms} and the references therein.}%
:
\begin{equation}
    \text{CVaR}_{p}(R) = \E{R|R\leq VaR_{p}(R)},
\label{eq:cvar}
\end{equation}
where $p$ denotes the confidence interval and the Value at Risk
VaR$_{p}(R)$ is the $p$-quantile of the trajectory reward given by:
\begin{align*}
    \text{VaR}_{p}(R) = \inf\{r \in \mathbb R : P(R\leq r) > p\},
\end{align*}
In particular, 
we make use of two $p$-quantiles for $p\in\cbra{0.1,0.9}$ to capture the two tails 
of the distribution of $R$ (see discussion in Section \ref{sSec:betasign}).

\vspace{-1em}

\subsection{On the sign and values of the parameter $\beta$}
\label{sSec:betasign}

The sign of the risk parameter $\beta$ determines the optimization problem that is being solved
according to \eqref{eq:rsrlkl}. 
%
%
Thus, in the simulated experiments of Sections \ref{sSec:cartpole} and \ref{sSec:acrobot}, 
it is expected that the risk-averse approach ($\beta $$<$$ 0$) 
reduces the variance (and CVaR$_p$ values for $p>0.5$) of the distribution of the total reward.
In addition, the risk-seeking approach ($\beta $$>$$ 0$) 
does not guarantee, but can also help reduce the variance 
(and CVaR$_p$ values for $p<0.5$) of the distribution of the total reward. 
Such a reduction can be 
indicative of a better suited learning behavior for the RL policies estimated 
by the proposed algorithm compared to the risk-neutral RL methods. 
Since in the risk-seeking (or ``optimistic'') case of $\beta $$>$$ 0$, 
emphasis is given on the right tail of the distribution of the total reward, 
convergence to policies with high average return can be accelerated under certain values of the hyper-parameters 
of the system and certain sequences of random exploratory actions.
%
The hyper-parameters that can affect this behavior include, for example, the learning rate of the actor and critic models, and random sequences that generate the exploratory actions.
The selection of the policies that yield the best performance among different 
runs (e.g. runs with different learning rates) 
is often adopted. 
In this case, 
the risk-seeking approach can also lead to better policies in terms of reduced variance.
%

%

\subsection{Inverted Pendulum (Cart-Pole)}
\label{sSec:cartpole}



%
\begin{figure*}[t]
\centering
\begin{subfigure}[b]{0.32\textwidth}
\centering
\includegraphics[trim=0 0 0 0,clip,width=\textwidth]{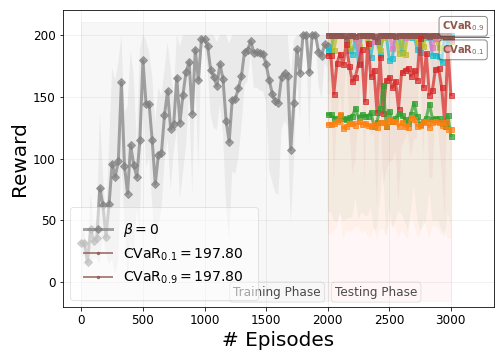}
\caption{Risk-neutral.}
\label{sfig:cartpole-tt-r-neutral}
\end{subfigure}
%
%
\begin{subfigure}[b]{0.32\textwidth}
\centering
\includegraphics[trim=0 0 0 0,clip,width=\textwidth]{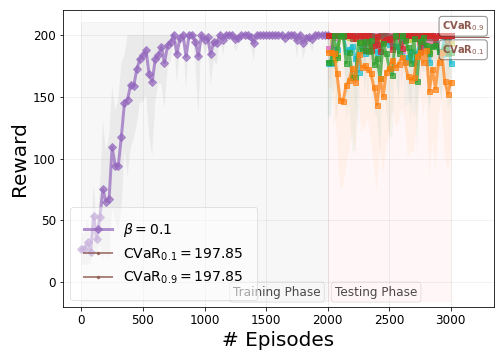}
\caption{Risk-seeking.}
\label{sfig:cartpole-tt-r-seeking}
\end{subfigure}
%
%
\begin{subfigure}[b]{0.32\textwidth}
\centering
\includegraphics[trim=0 0 0 0,clip,width=\textwidth]{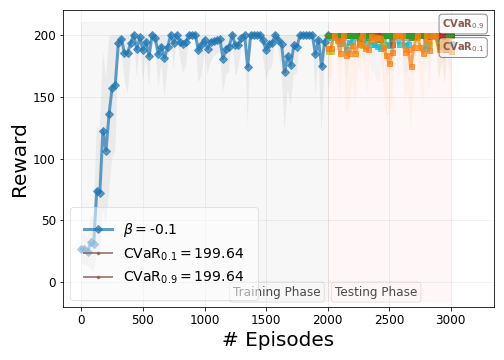}
\hspace{-1.4cm} \includegraphics[trim=0 -80 0 0,clip,width=0.2\textwidth]{legend}
\caption{Risk-averse.}
\label{sfig:cartpole-tt-r-averse}
\end{subfigure}
\begin{subfigure}[b]{0.1\textwidth}
\centering

\end{subfigure}
\caption{Training and testing behavior of the risk-neutral REINFORCE
algorithm against the proposed risk-sensitive R-REINFORCE algorithm 
(Alg. \ref{alg:RiskSensitiveREINFORCE})
for $\beta=-0.1$ and $\beta=+0.1$ in the Cart-Pole problem. 
Average reward, CVaR$_{0.1}$, and CVaR$_{0.9}$ values (for $l=0.5$) are computed over $10$ 
independent training and testing runs with different random seeds.}
\label{fig:cartpole-tt-r}
\end{figure*}
\begin{figure}[t]
\centering
\begin{subfigure}[b]{0.24\textwidth}
\centering
\includegraphics[trim=10 15 10 10,clip,width=\textwidth]{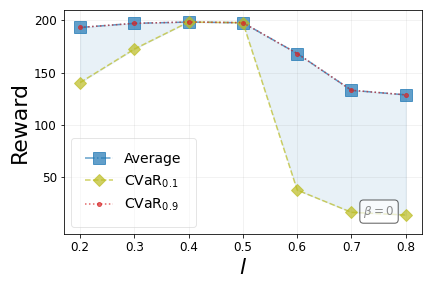}
\caption{$\beta = 0$.}
\label{sfig:cartpole-ll-r-zero}
\end{subfigure}
%
%
\begin{subfigure}[b]{0.24\textwidth}
\centering
\includegraphics[trim=10 15 10 10,clip,width=\textwidth]{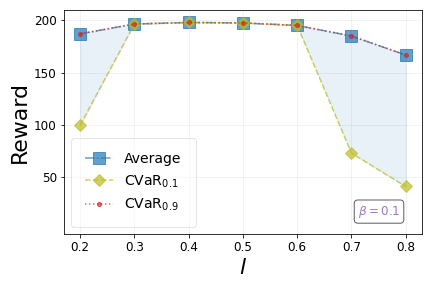}
\caption{$\beta = 0.1$.}
\label{sfig:cartpole-ll-r-01}
\end{subfigure}
\begin{subfigure}[b]{0.24\textwidth}
\centering
\includegraphics[trim=10 15 10 10,clip,width=\textwidth]{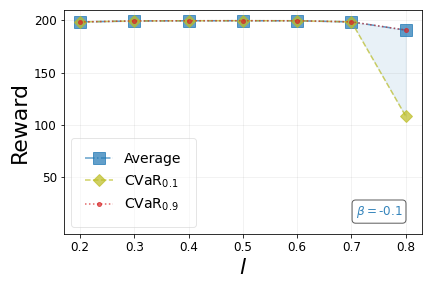}
\caption{$\beta = -0.1$.}
\label{sfig:cartpole-ll-r-m01}
\end{subfigure}
\caption{Robustness of
risk-neutral REINFORCE
and risk-sensitive R-REINFORCE 
(Alg. \ref{alg:RiskSensitiveREINFORCE}) algorithms in a cart-pole environment 
with respect to varying pole length. 
The training environment is modeled with pole length $l=0.5$. 
The testing environments have perturbed pole length values of $l\in\sbra{0.2,0.8}$.
Average reward, CVaR$_{0.1}$, and CVaR$_{0.9}$ values are computed over $10$ 
independent training and testing runs with different random seeds.}
\label{fig:cartpole-ll-r}
\vspace{-1.0em}
\end{figure}
%

In Figure \ref{fig:cartpole-tt-r} we present
the training and testing behavior of the risk-neutral REINFORCE
algorithm against the proposed risk-sensitive R-REINFORCE algorithm 
(Alg. \ref{alg:RiskSensitiveREINFORCE})
for $\beta=-0.1$ and $\beta=+0.1$ in the Cart-Pole problem.
We note that the use of REINFORCE with baseline
yielded no statistically significant results compared to risk-neutral REINFORCE.
%
The policy networks of the algorithms are modeled
as fully connected artificial neural networks with one hidden layer of 
only $h=16$ neurons and a `ReLU' activation function. 
The objective functions to be optimized are as defined in Section \ref{Sec:R-PG}.
We use a discount factor of $\gamma = 0.99$
and the `Adam' optimizer.
The best-performing learning rate within the set 
$\{0.001,0.003,0.005,0.007,0.01\}$ across all algorithms are selected for the visual inspection of the learning curves.
The algorithms are trained for $n_e=2000$ episodes 
in a training environment where the pole length is
$l=0.5$ and tested in different testing environments 
for $n_e=1000$ testing runs 
where the length of the pole
is perturbed such that $l\in\sbra{0.2,0.8}$.
The average reward for the different testing environments, as well as the 
CVaR$_{0.1}$, and CVaR$_{0.9}$ values for the testing environment without perturbations ($l=0.5$) are computed over $10$ 
independent training and testing runs with different random seeds.
%

We notice that 
although the mean, CVaR$_{0.1}$, and CVaR$_{0.9}$ metrics 
are not significantly different 
across the three algorithms,
the risk-sensitive algorithms in Fig. \ref{sfig:cartpole-tt-r-averse}
and Fig. \ref{sfig:cartpole-tt-r-seeking} converge faster to a near-optimal policy
that shows increased robustness with respect to model perturbations. 
This is further assessed in Fig. \ref{fig:cartpole-ll-r}, where 
the robustness of the algorithms with respect to model perturbations is quantified 
by the CVaR$_{0.1}$, and CVaR$_{0.9}$ values for all testing environments.
In Fig. \ref{sfig:cartpole-ll-r-zero}, we observe that the risk-neutral REINFORCE algorithm 
is performing very well near $l=0.5$, i.e., where no model perturbations exist,
but the performance is quickly deteriorated (CVaR$_{0.1}$ values decrease) in the 
presence of perturbations.
Fig. \ref{sfig:cartpole-ll-r-01} and Fig. \ref{sfig:cartpole-ll-r-m01} show that the 
risk-sensitive approaches increase the domain of perturbations where the behavior of
the RL agent is stable, with 
the risk-averse approach ($\beta<0$) showcasing the best behavior.


%
\begin{figure*}[t]
\centering
\begin{subfigure}[b]{0.32\textwidth}
\centering
\includegraphics[trim=0 0 0 0,clip,width=\textwidth]{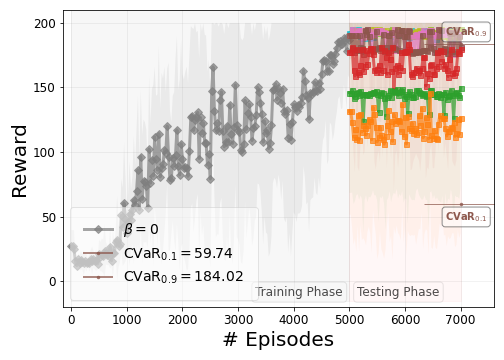}
\caption{Risk-neutral.}
\label{sfig:cartpole-tt-ac-neutral}
\end{subfigure}
%
%
\begin{subfigure}[b]{0.32\textwidth}
\centering
\includegraphics[trim=0 0 0 0,clip,width=\textwidth]{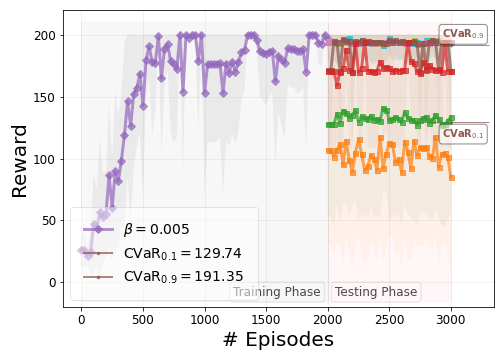}
\caption{Risk-seeking.}
\label{sfig:cartpole-tt-ac-seeking}
\end{subfigure}
%
%
\begin{subfigure}[b]{0.32\textwidth}
\centering
\includegraphics[trim=0 0 0 0,clip,width=\textwidth]{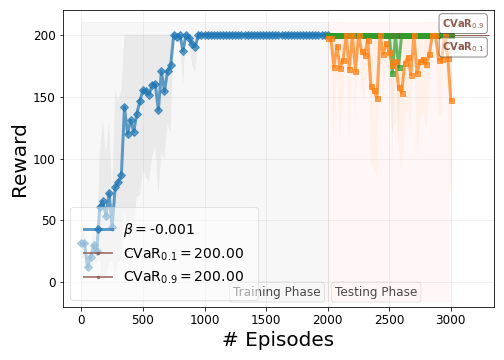}
\hspace{-1.4cm} \includegraphics[trim=0 -80 0 0,clip,width=0.2\textwidth]{legend}
\caption{Risk-averse.}
\label{sfig:cartpole-tt-ac-averse}
\end{subfigure}
%
%
%
\caption{Training and testing behavior of the risk-neutral Online Actor-Critic (OAC)
algorithm against the proposed risk-sensitive R-AC algorithm 
(Alg. \ref{alg:RiskSensitiveActorCritic})
for $\beta=-0.001$ and $\beta=+0.005$ in the Cart-Pole problem. 
Average reward, CVaR$_{0.1}$, and CVaR$_{0.9}$ values (for $l=0.5$) are computed over $10$ 
independent training and testing runs with different random seeds.}
\label{fig:cartpole-tt-ac}
\end{figure*}
\begin{figure}[t]
\centering
\begin{subfigure}[b]{0.24\textwidth}
\centering
\includegraphics[trim=10 15 10 10,clip,width=\textwidth]{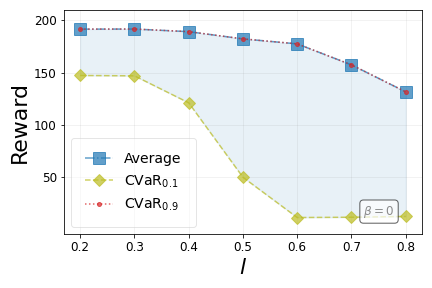}
\caption{$\beta = 0$.}
\label{sfig:cartpole-ll-ac-zero}
\end{subfigure}
%
%
\begin{subfigure}[b]{0.24\textwidth}
\centering
\includegraphics[trim=10 15 10 10,clip,width=\textwidth]{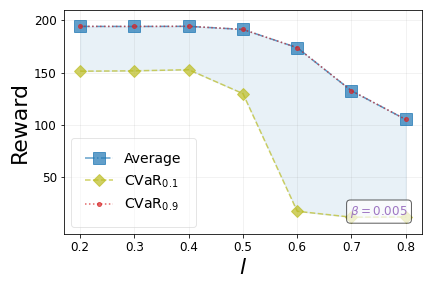}
\caption{$\beta = 0.005$.}
\label{sfig:cartpole-ll-ac-005}
\end{subfigure}
%
%
\begin{subfigure}[b]{0.24\textwidth}
\centering
\includegraphics[trim=10 15 10 10,clip,width=\textwidth]{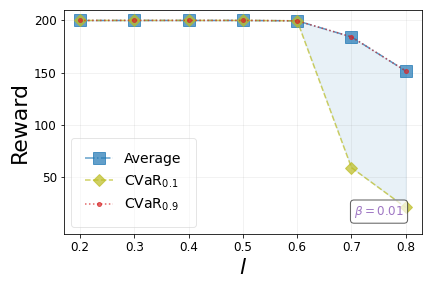}
\caption{$\beta = 0.01$.}
\label{sfig:cartpole-ll-ac-01}
\end{subfigure}
%
%
\begin{subfigure}[b]{0.24\textwidth}
\centering
\includegraphics[trim=10 15 10 10,clip,width=\textwidth]{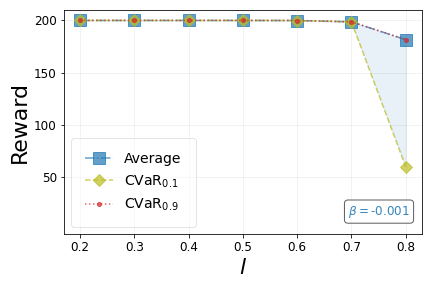}
\caption{$\beta = -0.001$.}
\label{sfig:cartpole-ll-ac-m001}
\end{subfigure}
\begin{subfigure}[b]{0.24\textwidth}
\centering
\includegraphics[trim=10 15 10 10,clip,width=\textwidth]{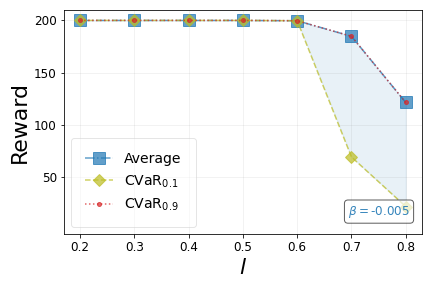}
\caption{$\beta = -0.005$.}
\label{sfig:cartpole-ll-ac-m005}
\end{subfigure}
\begin{subfigure}[b]{0.24\textwidth}
\centering
\includegraphics[trim=10 15 10 10,clip,width=\textwidth]{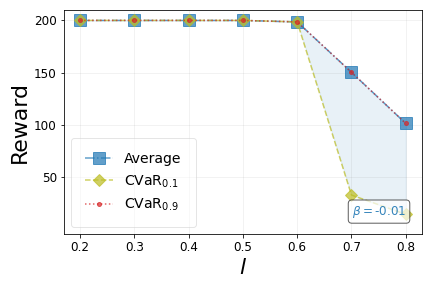}
\caption{$\beta = -0.01$.}
\label{sfig:cartpole-ll-ac-m01}
\end{subfigure}
\caption{Robustness of
risk-neutral Online Actor-Critic (OAC)
and risk-sensitive R-AC 
(Alg. \ref{alg:RiskSensitiveActorCritic}) algorithms in a cart-pole environment 
with respect to varying pole length. 
The training environment is modeled with pole length $l=0.5$. 
The testing environments have perturbed pole length values of $l\in\sbra{0.2,0.8}$.
Average reward, CVaR$_{0.1}$, and CVaR$_{0.9}$ values are computed over $10$ 
independent training and testing runs with different random seeds.}
\label{fig:cartpole-ll-ac}
\vspace{-1.0em}
\end{figure}
\begin{figure}[t]
\centering
\begin{subfigure}[b]{0.24\textwidth}
\centering
\includegraphics[trim=0 0 0 0,clip,width=\textwidth]{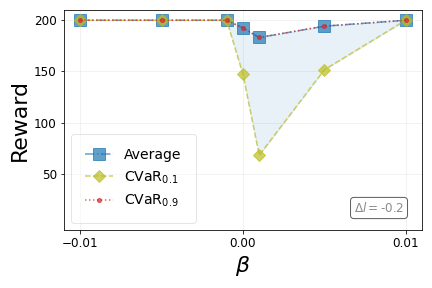}
\caption{$l=0.3$.}
\label{sfig:cartpole-bb-ac-minus}
\end{subfigure}
%
%
\begin{subfigure}[b]{0.24\textwidth}
\centering
\includegraphics[trim=0 0 0 0,clip,width=\textwidth]{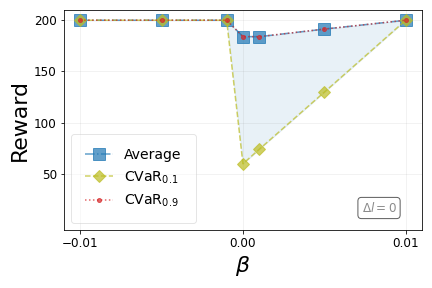}
\caption{$l=0.5$.}
\label{sfig:cartpole-bb-ac-zero}
\end{subfigure}
%
%
\begin{subfigure}[b]{0.24\textwidth}
\centering
\includegraphics[trim=0 0 0 0,clip,width=\textwidth]{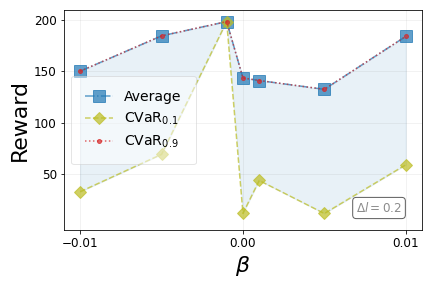}
\caption{$l=0.7$.}
\label{sfig:cartpole-bb-ac-plus}
\end{subfigure}
\caption{Sensitivity analysis of the risk-sensitive R-AC algorithm 
(Alg. \ref{alg:RiskSensitiveActorCritic})
with respect to the risk-sensitive parameter $\beta\in\sbra{-0.01,0.01}$ in the Cart-Pole problem. 
$\beta=0$ corresponds to the risk-neutral Online Actor-Critic (OAC).
The training environment is modeled with pole length $l=0.5$. 
Average reward, CVaR$_{0.1}$, and CVaR$_{0.9}$ values for testing environments with 
$l\in\{0.3,0.5,0.7\}$ are computed over $10$ 
independent training and testing runs with different random seeds.}
\label{fig:cartpole-bb-ac}
\vspace{-1.0em}
\end{figure}
%

In Figure \ref{fig:cartpole-tt-ac} we present
the training and testing behavior of the 
risk-neutral Online Actor-Critic (OAC)
and risk-sensitive actor critic (R-AC) 
(Alg. \ref{alg:RiskSensitiveActorCritic}) algorithms in the cart-pole environment 
with respect to varying pole length. 
The policy networks of the algorithms are modeled
as fully connected artificial neural networks with one hidden layer of 
only $h=16$ neurons and a `ReLU' activation function. 
The objective functions to be optimized are as defined in Section \ref{sSec:RSactorcritic}.
We use a discount factor of $\gamma = 0.99$
and the `Adam' optimizer with the best-performing learning rates within the set 
$\{0.0003,0.0005,0.0007,0.001\}$ across all algorithms. 
The best-performing learning rate are chosen best on visual inspection of the learning curves.
The algorithms are trained for $n_e=2000$ episodes 
in a training environment where the pole length is
$l=0.5$ and tested in different testing environments 
for $n_e=1000$ testing runs 
where the length of the pole
is perturbed such that $l\in\sbra{0.2,0.8}$.
The average reward for the different testing environments, as well as the 
CVaR$_{0.1}$, and CVaR$_{0.9}$ values for the testing environment without perturbations ($l=0.5$) are computed over $10$ 
independent training and testing runs with different random seeds.
%

We notice that 
although the mean value performance 
is not significantly different 
across the three algorithms,
the risk-sensitive algorithms in Fig. \ref{sfig:cartpole-tt-ac-averse}
and Fig. \ref{sfig:cartpole-tt-ac-seeking} converge to a near-optimal policy
(in the risk-averse case the performance is optimal)
that shows reduced variation across different runs, as indicated 
by the CVaR$_{0.1}$, and CVaR$_{0.9}$ values calculated for $l=0.5$ (no model perturbations). 
Moreover, notice that the risk-neutral algorithm in 
\ref{sfig:cartpole-tt-ac-neutral} is trained for $n_e=5000$ episodes to achieve 
similar performance to the risk-sensitive algorithms. 
This indicates better sample efficiency for the proposed risk-sensitive algorithms 
in Alg. \ref{alg:RiskSensitiveActorCritic}.
The robustness of the algorithms with respect to model perturbation 
is further assessed in Fig. \ref{fig:cartpole-ll-ac}.
Fig. \ref{sfig:cartpole-ll-ac-zero}, shows how the CVaR$_{0.1}$ values decrease 
as the pole length increases in the risk-neutral case ($\beta=0$).
Fig. \ref{sfig:cartpole-ll-ac-005} and Fig. \ref{sfig:cartpole-ll-ac-01} show that the 
risk-seeking approaches slightly increase the robustness of the learned policies. 
However, as shown in 
Fig. \ref{sfig:cartpole-ll-ac-m001}, Fig. \ref{sfig:cartpole-ll-ac-m005}, and 
Fig. \ref{sfig:cartpole-ll-ac-m01},
the risk-averse approach ($\beta<0$) showcases significantly increased robustness
with respect to perturbations in the pole length. 

%
Fig. \ref{fig:cartpole-bb-ac} presents a sensitivity analysis of the 
algorithms with respect to the risk-sensitive parameter $\beta\in\sbra{-0.01,0.01}$. 
Three testing environments are studied for 
$l=0.5$ (no perturbation),
$l=0.3$ (overestimation during training), and 
$l=0.7$ (underestimation during training).
Negative values for $\beta$ showcase a more stable behavior across the
testing environments.
Moreover, notice that $sgn(\beta)<0$ is roughly adequate for 
a stable behavior regardless of the numerical value of $\beta$, 
as long as it is close to zero, i.e., 
no precise estimation of the optimal $\beta$ is required.

\subsection{Underactuated Double Pendulum (Acrobot)}
\label{sSec:acrobot}



%
\begin{figure}[t]
\centering
\begin{subfigure}[b]{0.24\textwidth}
\centering
\includegraphics[trim=0 0 0 0,clip,width=\textwidth]{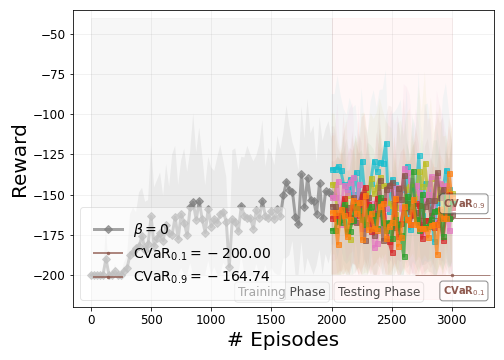}
\hspace{-1.1cm} \includegraphics[trim=0 -180 0 0,clip,width=0.2\textwidth]{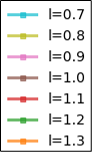}
\caption{Risk-neutral.}
\label{sfig:acrobot-tt-r-neutral}
\end{subfigure}
\begin{subfigure}[b]{0.24\textwidth}
\centering
\includegraphics[trim=0 0 0 0,clip,width=\textwidth]{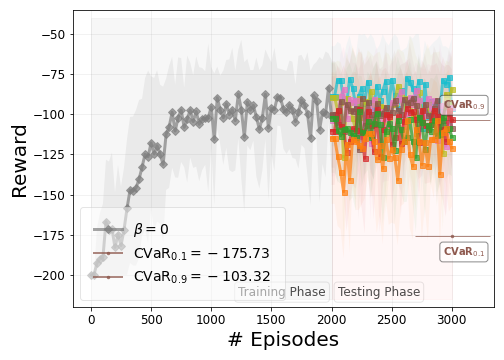}
\caption{Risk-neutral with Baseline.}
\label{sfig:acrobot-tt-r-neutral-BL}
\end{subfigure}
%
%
%
\begin{subfigure}[b]{0.24\textwidth}
\centering
\includegraphics[trim=0 0 0 0,clip,width=\textwidth]{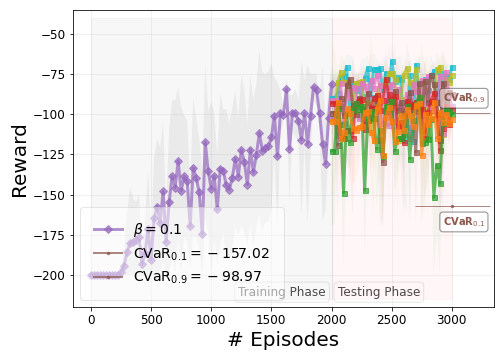}
\caption{Risk-seeking.}
\label{sfig:acrobot-tt-r-seeking}
\end{subfigure}
%
%
\begin{subfigure}[b]{0.24\textwidth}
\centering
\includegraphics[trim=0 0 0 0,clip,width=\textwidth]{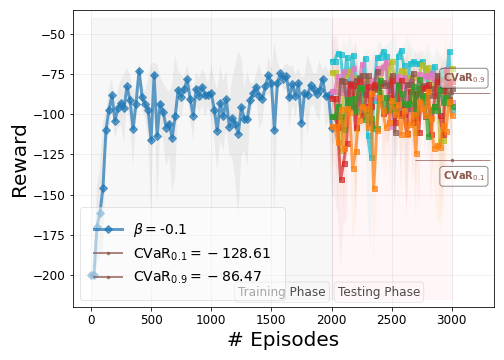}
\caption{Risk-averse.}
\label{sfig:acrobot-tt-r-averse}
\end{subfigure}
\caption{Training and testing behavior of the risk-neutral REINFORCE
and 
risk-neutral REINFORCE with baseline
algorithms against the proposed risk-sensitive R-REINFORCE algorithm 
(Alg. \ref{alg:RiskSensitiveREINFORCE})
for $\beta=-0.1$ and $\beta=+0.1$ in the Acrobot problem. 
Average reward, CVaR$_{0.1}$, and CVaR$_{0.9}$ values (for $l=1.0$) are computed over $10$ 
independent training and testing runs with different random seeds.}
\label{fig:acrobot-tt-r}
\vspace{-1.0em}
\end{figure}
\begin{figure}[t]
\centering
\begin{subfigure}[b]{0.23\textwidth}
\centering
\includegraphics[trim=0 0 0 0,clip,width=\textwidth]{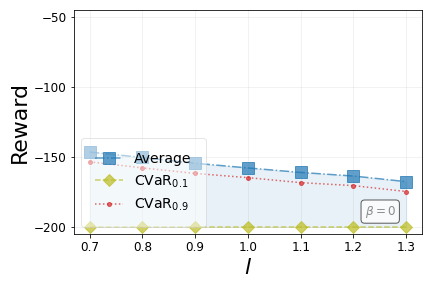}
\caption{$\beta = 0$.}
\label{sfig:acrobot-ll-r-zero}
\end{subfigure}
\begin{subfigure}[b]{0.23\textwidth}
\centering
\includegraphics[trim=0 0 0 0,clip,width=\textwidth]{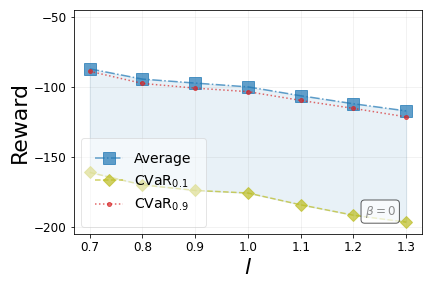}
\caption{$\beta = 0$ with Baseline.}
\label{sfig:acrobot-ll-r-zero-BL}
\end{subfigure}
\begin{subfigure}[b]{0.23\textwidth}
\centering
\includegraphics[trim=0 0 0 0,clip,width=\textwidth]{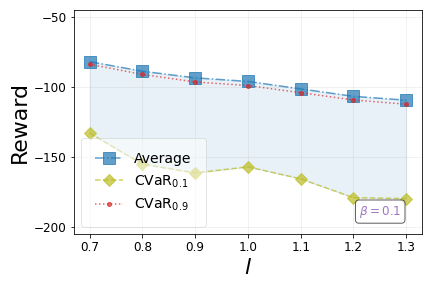}
\caption{$\beta = 0.005$.}
\label{sfig:acrobot-ll-r-01}
\end{subfigure}
\begin{subfigure}[b]{0.23\textwidth}
\centering
\includegraphics[trim=0 0 0 0,clip,width=\textwidth]{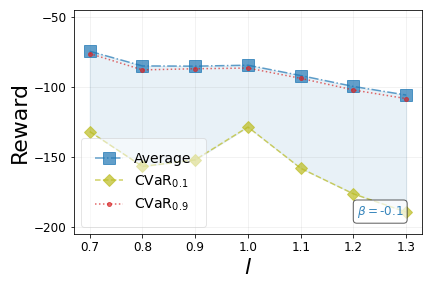}
\caption{$\beta = -0.001$.}
\label{sfig:acrobot-ll-r-m01}
\end{subfigure}
\caption{Robustness of
risk-neutral REINFORCE
risk-neutral REINFORCE with baseline,
and risk-sensitive R-REINFORCE 
in the Acrobot environment 
with respect to varying pole length. 
The training environment is modeled with pole length $l=1.0$. 
The testing environments have perturbed pole length values of $l\in\sbra{0.7,1.3}$.
Average reward, CVaR$_{0.1}$, and CVaR$_{0.9}$ values are computed over $10$ 
independent training and testing runs with different random seeds.}
\label{fig:acrobot-ll-r}
\vspace{-1.0em}
\end{figure}
%

In Figure \ref{fig:acrobot-tt-r} we present
the training and testing behavior of the risk-neutral REINFORCE
with and without baseline 
algorithms against the proposed risk-sensitive R-REINFORCE algorithm 
(Alg. \ref{alg:RiskSensitiveREINFORCE})
for $\beta=-0.1$ and $\beta=+0.1$ in the Acrobot problem.
The policy networks of the algorithms are modeled
as fully connected artificial neural networks with one hidden layer of 
only $h=64$ neurons and a `ReLU' activation function. 
The objective functions to be optimized are as defined in Section \ref{Sec:R-PG}.
We use a discount factor of $\gamma = 0.99$
and the `Adam' optimizer with the best performing learning rates within the set 
$\{0.001,0.003,0.005,0.007,0.01\}$ across all algorithms.
The algorithms are trained for $n_e=2000$ episodes 
in a training environment where the pole length of the first link is
$l=1.0$ and tested in different testing environments 
for $n_e=1000$ testing runs 
where the length of the pole
is perturbed such that $l\in\sbra{0.7,1.3}$.
The average reward for the different testing environments, as well as the 
CVaR$_{0.1}$, and CVaR$_{0.9}$ values for the testing environment without perturbations ($l=1.0$) are computed over $10$ 
independent training and testing runs with different random seeds.
%

First, we notice (Fig. \ref{sfig:acrobot-tt-r-neutral-BL}) that risk-neutral REINFORCE without baseline is not able to 
learn a policy that solves the Acrobot problem. 
On the remaining algorithms,  
although the mean performance is not significantly different,
the risk-sensitive algorithms in Fig. \ref{sfig:acrobot-tt-r-averse}
and Fig. \ref{sfig:acrobot-tt-r-seeking} showcase 
increased CVaR$_{0.1}$ values that suggest reduced variation across different runs.
The fact that the risk-sensitive approaches perform on par, and slightly better,
compared to REINFORCE with baseline, is indicative of the 
implicit baseline term present in optimizing for the exponential objective function
as explained in Section \ref{sSec:preliminaries-exp}.
The robustness of the algorithms with respect to model perturbation 
is further assessed in Fig. \ref{fig:acrobot-ll-r} for all testing environments.
Similar to the Cart-Pole problem, 
Fig. \ref{fig:acrobot-ll-r} suggests that the 
risk-sensitive approaches can increase the domain of perturbations where the behavior of
the RL agent is more stable, with 
the risk-averse approach ($\beta<0$) showcasing the best behavior. 
%


%
\begin{figure}[t]
\centering
\begin{subfigure}[b]{0.24\textwidth}
\centering
\includegraphics[trim=0 0 0 0,clip,width=\textwidth]{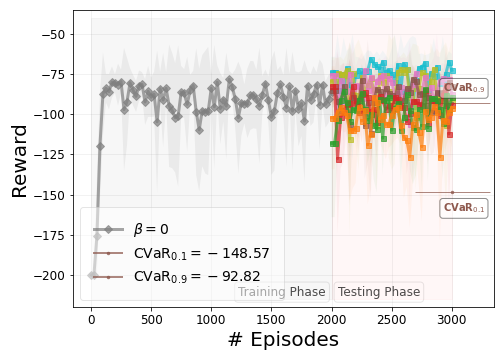}
\caption{Risk-neutral.}
\label{sfig:acrobot-tt-ac-neutral}
\end{subfigure}
%
%
\begin{subfigure}[b]{0.24\textwidth}
\centering
\includegraphics[trim=0 0 0 0,clip,width=\textwidth]{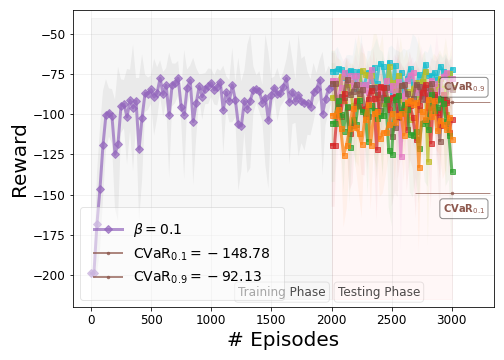}
\caption{Risk-seeking.}
\label{sfig:acrobot-tt-ac-seeking}
\end{subfigure}
%
%
\begin{subfigure}[b]{0.24\textwidth}
\centering
\includegraphics[trim=0 0 0 0,clip,width=\textwidth]{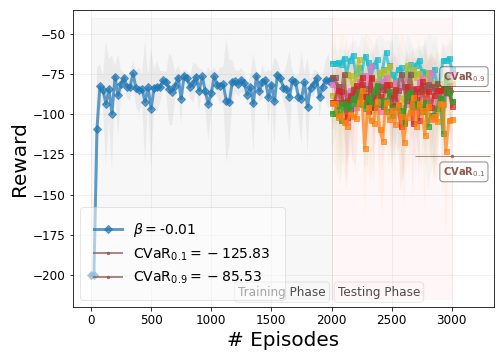}
\caption{Risk-averse.}
\label{sfig:acrobot-tt-ac-averse}
\end{subfigure}
\begin{subfigure}[b]{0.1\textwidth}
\centering
\includegraphics[trim=0 -110 0 0,clip,width=0.6\textwidth]{legend-a}
\end{subfigure}
\caption{Training and testing behavior of the risk-neutral Online Actor-Critic 
algorithm against the proposed risk-sensitive R-AC algorithm 
(Alg. \ref{alg:RiskSensitiveActorCritic})
for $\beta=-0.01$ and $\beta=+0.1$ in the Acrobot problem. 
Average reward, CVaR$_{0.1}$, and CVaR$_{0.9}$ values (for $l=1.0$) are computed over $10$ 
independent training and testing runs with different random seeds.}
\label{fig:acrobot-tt-ac}
\vspace{-1.0em}
\end{figure}
\begin{figure}[t]
\centering
\begin{subfigure}[b]{0.24\textwidth}
\centering
\includegraphics[trim=0 0 0 0,clip,width=\textwidth]{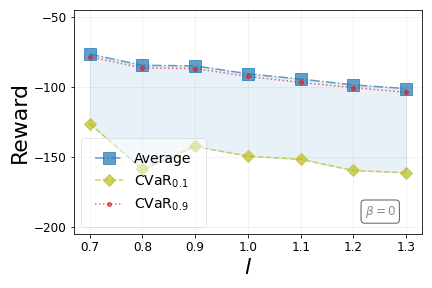}
\caption{$\beta = 0$.}
\label{sfig:acrobot-ll-ac-zero}
\end{subfigure}
%
%
\begin{subfigure}[b]{0.24\textwidth}
\centering
\includegraphics[trim=0 0 0 0,clip,width=\textwidth]{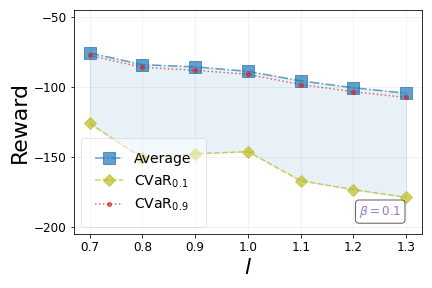}
\caption{$\beta = 0.005$.}
\label{sfig:acrobot-ll-ac-005}
\end{subfigure}
\begin{subfigure}[b]{0.24\textwidth}
\centering
\includegraphics[trim=0 0 0 0,clip,width=\textwidth]{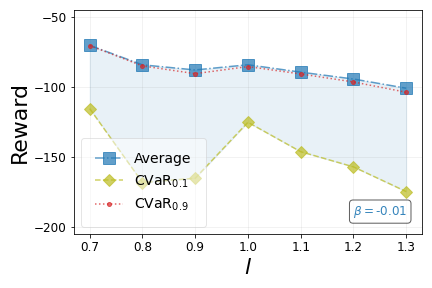}
\caption{$\beta = -0.001$.}
\label{sfig:acrobot-ll-ac-m001}
\end{subfigure}
\caption{Robustness of
risk-neutral Online Actor-Critic (OAC)
and risk-sensitive R-AC 
(Alg. \ref{alg:RiskSensitiveActorCritic}) algorithms in the Acrobot environment 
with respect to varying pole length. 
The training environment is modeled with pole length $l=1.0$. 
The testing environments have perturbed pole length values of $l\in\sbra{0.7,1.3}$.
Average reward, CVaR$_{0.1}$, and CVaR$_{0.9}$ values are computed over $10$ 
independent training and testing runs with different random seeds.}
\label{fig:acrobot-ll-ac}
\vspace{-1.0em}
\end{figure}
%

Finally, in Figure \ref{fig:acrobot-tt-ac} we present
the training and testing behavior of the 
risk-neutral Online Actor-Critic (OAC)
and risk-sensitive actor critic (R-AC) 
(Alg. \ref{alg:RiskSensitiveActorCritic}) algorithms in the Acrobot environment 
with respect to varying pole length. 
The policy networks of the algorithms are modeled
as fully connected artificial neural networks with one hidden layer of 
only $h=64$ neurons and a `ReLU' activation function. 
The objective functions to be optimized are as defined in Section \ref{sSec:RSactorcritic}.
We use a discount factor of $\gamma = 0.99$
and the `Adam' optimizer with the best performing learning rates within the set 
$\{0.0003,0.0005,0.0007,0.001\}$ across all algorithms.
The algorithms are trained for $n_e=2000$ episodes 
in a training environment where the pole length of the first link is
$l=1.0$ and tested in different testing environments 
for $n_e=1000$ testing runs 
where the length of the pole
is perturbed such that $l\in\sbra{0.7,1.3}$.
The average reward for the different testing environments, as well as the 
CVaR$_{0.1}$, and CVaR$_{0.9}$ values for the testing environment without perturbations ($l=1.0$) are computed over $10$ 
independent training and testing runs with different random seeds.
%

Similar to the Cart-Pole case, we notice that 
although the mean value performance 
is not significantly different 
across the three algorithms,
the risk-sensitive algorithms in Fig. \ref{sfig:acrobot-tt-ac-averse}
and Fig. \ref{sfig:acrobot-tt-ac-seeking} converge to a near-optimal policy
that shows reduced variation across different runs, as indicated 
by the CVaR$_{0.1}$, and CVaR$_{0.9}$ values calculated for $l=1.0$ (no model perturbations). 
The robustness of the algorithms with respect to model perturbation 
is further assessed in Fig. \ref{fig:acrobot-ll-ac}.
Fig. \ref{sfig:acrobot-ll-ac-zero}, shows how the CVaR$_{0.1}$ values decrease 
as the pole length increases in the risk-neutral case ($\beta=0$).
Fig. \ref{sfig:acrobot-ll-ac-m001} shows that the 
risk-averse approach can increase the robustness of the learned policies for small
perturbations. 

These results are consistent with the analysis presented in Section \ref{sec:preliminaries} 
and suggest that the use of exponential criteria results 
can inherit 
computational and convergence properties of standard RL algorithms, 
while accelerating the learning process, leading in increased sample efficiency, 
and policies with increased robustness with respect to 
environmental and model perturbations.

\section{Conclusion}
\label{Sec:Conclusion}

Risk-sensitive reinforcement learning algorithms are receiving increasing attention
in an endeavor to construct 
robust and sample-efficient algorithms which can lead to increased performance
compared to their risk-neutral counterparts.
We formulate a risk-sensitive reinforcement learning approach as
an optimization problem with respect to a modified objective based on 
exponential criteria.
In particular, 
we study a model-free risk-sensitive variation of the widely-used 
Monte Carlo Policy Gradient algorithm, and introduce a novel
risk-sensitive online Actor-Critic algorithm based on 
solving a multiplicative Bellman equation using stochastic approximation updates.
Analytical results suggest that the use of exponential criteria 
generalizes commonly used ad-hoc regularization approaches, 
improves sample efficiency, and 
introduces robustness with respect to perturbations in the model parameters and 
the environment.
The implementation, performance, and robustness properties of the proposed methods
are evaluated in simulated experiments, and suggest suitability for real-life 
applications where learning in simulation is followed by transferring the learned policies
to real agents in noisy environments.

\bibliographystyle{IEEEtran} %
\bibliography{bib_rsrl.bib}


\appendices

\section{Risk-Sensitive Policy Gradient Update Rule} 
\label{apndx:derivation}

In this section, we provide a risk-sensitive version of the policy gradient theorem \cite{sutton1999policy} using exponential criteria, which is used to derive 
the update rule for the Risk-Sensitive REINFORCE algorithm in \eqref{update_rule_exp}. 
The exponential objective can be written as an integral (summation for finite state and action spaces) over all possible trajectories, i.e.,
\begin{align} \label{grad1}
    \nabla_{\theta} J_{\beta}(\theta) &= \nabla \frac{1}{\beta} \int \rho_{\theta}(\tau) \exp \{\beta R(\tau)\} d\tau \notag\\
    &= \frac{1}{\beta} \int \rho_{\theta}(\tau) \frac{\nabla \rho_{\theta}(\tau)}{\rho_{\theta}(\tau)} \exp \{\beta R(\tau)\} d\tau \notag\\
    &= \frac{1}{\beta} \int \rho_{\theta}(\tau) \nabla \log \rho_{\theta}(\tau) \exp \{\beta R(\tau)\} d\tau \notag\\
    &= \frac{1}{\beta} \mathbb{E}_{\tau \sim \rho_{\theta}} \Big[ \nabla \log \rho_{\theta}(\tau) \exp \{\beta R(\tau)\} \Big]
\end{align}
Using the ``log-trick'' \cite{sutton2018reinforcement}, the gradient of the $J_{\beta}(\theta)$ with respect to the policy parameter $\theta$ can be obtained as follows,
\begin{align}
    \nabla_{\theta} J_{\beta}(\theta) &= \nabla \frac{1}{\beta} \int \rho_{\theta}(\tau) \exp \{\beta R(\tau)\} d\tau \notag\\
    &= \frac{1}{\beta} \mathbb{E}_{\tau \sim \rho_{\theta}} \Big[ \nabla \log \rho_{\theta}(\tau) \exp \{\beta R(\tau)\} \Big]
\end{align}
Recall that $\rho_{\theta}(\tau) $$=$$ p_0 \prod_{t=0}^{|\tau|-1} \pi(a_t|s_{t}; \theta) p(s_{t+1}|s_t,a_t)$. 
Then, by first taking the logarithm and then the gradient of both sides, we get
\begin{equation} 
    \begin{aligned} \label{grad-traj-dist}
    \nabla \log \rho_{\theta}(\tau) = \sum_{t=0}^{|\tau|-1} \nabla \log \pi(a_t|s_t;\theta)
    \end{aligned}
\end{equation}
For brevity, we use $\pi_t(\theta) $$:=$$ \pi(a_t|s_t;\theta)$. Thus, by substituting Eq. \eqref{grad-traj-dist} in Eq. \eqref{grad1}, we get
\begin{equation} 
    \begin{aligned}
        \nabla J_{\theta}(\theta) = \frac{1}{\beta} \mathbb{E}_{\tau \sim \rho_{\theta}} \Big[ \sum_{t=0}^{|\tau|-1} \nabla \log \pi_t(\theta) \exp\{\beta R(\tau)\} \Big]
    \end{aligned}
\end{equation}
Recall that $R(\tau) = \sum_{t=0}^{|\tau|-1} \gamma^t r(s_t, a_t)$. Using this fact and the property of exponential, we have
\begin{equation} 
    \begin{aligned}
        \nabla J_{\theta}(\theta) = \frac{1}{\beta} \mathbb{E}_{\tau \sim \rho_{\theta}} \Big[ \sum_{t=0}^{|\tau|-1} & \nabla \log \pi_t(\theta)
        \exp\{\beta \sum_{t'=0}^{t-1} \gamma^{t'} r(s_t, a_t)\} 
        \\& 
        \exp\{\beta \sum_{t'=t}^{|\tau|-1} \gamma^{t'} r(s_{t'}, a_{t'})\} \Big]
    \end{aligned}
\end{equation}
By using the temporal structure of the problem and causality, it can be argued that the rewards prior to time $t$ are not dependent on the actions that the policy will take in a future state $s_t$, that is, $\sum_{t'=0}^{t-1} \gamma^{t'} r_t(s_{t'},a_{t'})$ is independent of $\nabla \log \pi(a_t|s_t;\theta)$. Thus, by using the independence property, we have
\begin{equation} 
    \begin{aligned} \label{adj-sep}
        &\nabla J_{\theta}(\theta) = \frac{1}{\beta} \mathbb{E}_{\tau \sim \rho_{\theta}} \Big[\exp \{\beta \sum_{t'=0}^{t-1} \gamma^{t'} r(s_{t'}, a_{t'})\} \Big] 
        \\ & 
        \quad \cdot 
        \mathbb{E}_{\tau \sim \rho_{\theta}} \Big[ \sum_{t=0}^{|\tau|-1} \nabla \log \pi_t(\theta) 
        \exp \{\beta \sum_{t'=t}^{|\tau|-1} \gamma^{t'} r(s_{t'}, a_{t'})\} \Big]
    \end{aligned}
\end{equation}
Note that the first expectation is a constant, therefore,
\begin{equation} 
    \begin{aligned}
        \nabla J_{\theta}(\theta) \propto \mathbb{E}_{\tau \sim \rho_{\theta}} \Big[ \sum_{t=0}^{|\tau|-1} \frac{1}{\beta} e^{\beta R_t} \nabla \log \pi_t(\theta)\Big] \\
    \end{aligned}
\end{equation}
where $R_t $$:=$$ \sum_{t'=t}^{|\tau|-1} \gamma^{t'-t} r(s_{t'}, a_{t'})$. 

As a final remark, notice that from \eqref{adj-sep}, we can see that the first term on the right-hand side of the equation provides an inherent way of adjusting the step size, effectively making the constant step size adaptive.

\subsection{Convergence Analysis} \label{apndx:convergence}

In this section we show that the the parameter vector $\theta$ updated by the 
risk-sensitive REINFORCE algorithm in \eqref{update_rule_exp}
converges to the optimal parameter vector $\theta^*$ 
in expectation, 
for sufficiently small values of the risk-parameter $\beta$. 
First note the following identity
\begin{align*}
    \| \theta_{t+1} &- \theta^* \|^2 -  \| \theta_t - \theta^* \|^2 \\
    &= \| \theta_{t+1} -\theta_t + \theta_t - \theta^* \|^2 -  \| \theta_t - \theta^* \|^2 \\
    &= \| \theta_{t+1} - \theta_t \|^2 -2(\theta_{t+1} - \theta_t) \cdot (\theta_t - \theta^*)
\end{align*}    
Using the R-REINFORCE update rule in \eqref{update_rule_exp}, i.e., 
\begin{equation*}
\theta_{t+1} = \theta_t +\frac{\eta }{\beta} e^{\beta R_{t}^+}  \nabla \log \pi_{\theta_t}(a_t|s_t)    
\end{equation*}
we get
\begin{align*}
    \| \theta_{t+1} - \theta^* \|^2 -  \| \theta_t &- \theta^* \|^2 = (\frac{\eta}{\beta} e^{\beta R_{t}^+})^2 \| \nabla \log \pi_{\theta_t}(a_t|s_t) \|^2 \\ & - 2  \frac{\eta}{\beta} e^{\beta R_{t}^+}  \nabla \log \pi_{\theta_t}(a_t|s_t) \cdot (\theta_t - \theta^*)
\end{align*}
By taking the conditional expectation with filtration $\mathcal F_t$ from both sides of the equation, we have
\begin{align*}
    \mathbb{E} &\Big[ \| \theta_{t+1} - \theta^* \|^2\mid \mathcal F_t \Big] = \| \theta_t - \theta^* \|^2 \\ & \qquad + (\frac{\eta}{\beta})^2 \mathbb{E} \Big[ e^{2\beta R_{t}^+} \| \nabla \log \pi_{\theta_t}(a_t|s_t) \|^2 \mid \mathcal F_t \Big]\\ &  \qquad - 2 \frac{\eta}{\beta} e^{-\beta R_{t}^-} \mathbb{E} \Big[ e^{\beta R}  \nabla \log \pi_{\theta_t}(a_t|s_t) \mid \mathcal F_t] \cdot (\theta_t - \theta^*) \\
    &= \| \theta_t - \theta^* \|^2 + (\frac{\eta}{\beta})^2 \mathbb{E} \Big[ e^{2\beta R_{t}^+} \| \nabla \log \pi_{\theta_t}(a_t|s_t) \|^2 \mid \mathcal F_t \Big] \\ & \qquad - 2 \eta e^{-\beta R_{t}^-} \nabla J_{\gamma}(\theta_t) \cdot (\theta_t - \theta^*)
\end{align*}
The first line follows from the conditioning on the filtration $\mathcal F_t$. The second line follows from the fact that $\nabla J_{\gamma}(\theta_t) = \mathbb{E} \Big[\frac{1}{\beta} e^{\beta R}  \nabla \log \pi_{\theta_t}(a_t|s_t) \mid \mathcal F_t \Big]$. 
It should be noted that since $\theta^* = argmax_{\theta} J_{\gamma}(\theta)$, it follows that 
$\nabla J_{\gamma}(\theta_t) \cdot (\theta_t - \theta^*) $$>$$ 0$. 
Finally, it follows that $\theta_t$ converges to $\theta^*$, as long as the following condition holds:
\begin{align*}
    (\frac{\eta}{\beta})^2 \mathbb{E} \Big[ e^{2\beta R_{t}^+} &\| \nabla \log \pi_{\theta_t}(a_t|s_t) \|^2 \mid \mathcal F_t \Big] \\ &- 2 \eta e^{-\beta R_{t}^+} \nabla J_{\gamma}(\theta_t) \cdot (\theta_t - \theta^*) < 0.
\end{align*}


\vskip 0.2in

\begin{IEEEbiography}[{\includegraphics[width=1in,height=1.25in,clip, trim= 6cm 0.0cm 7cm 0.0cm,]{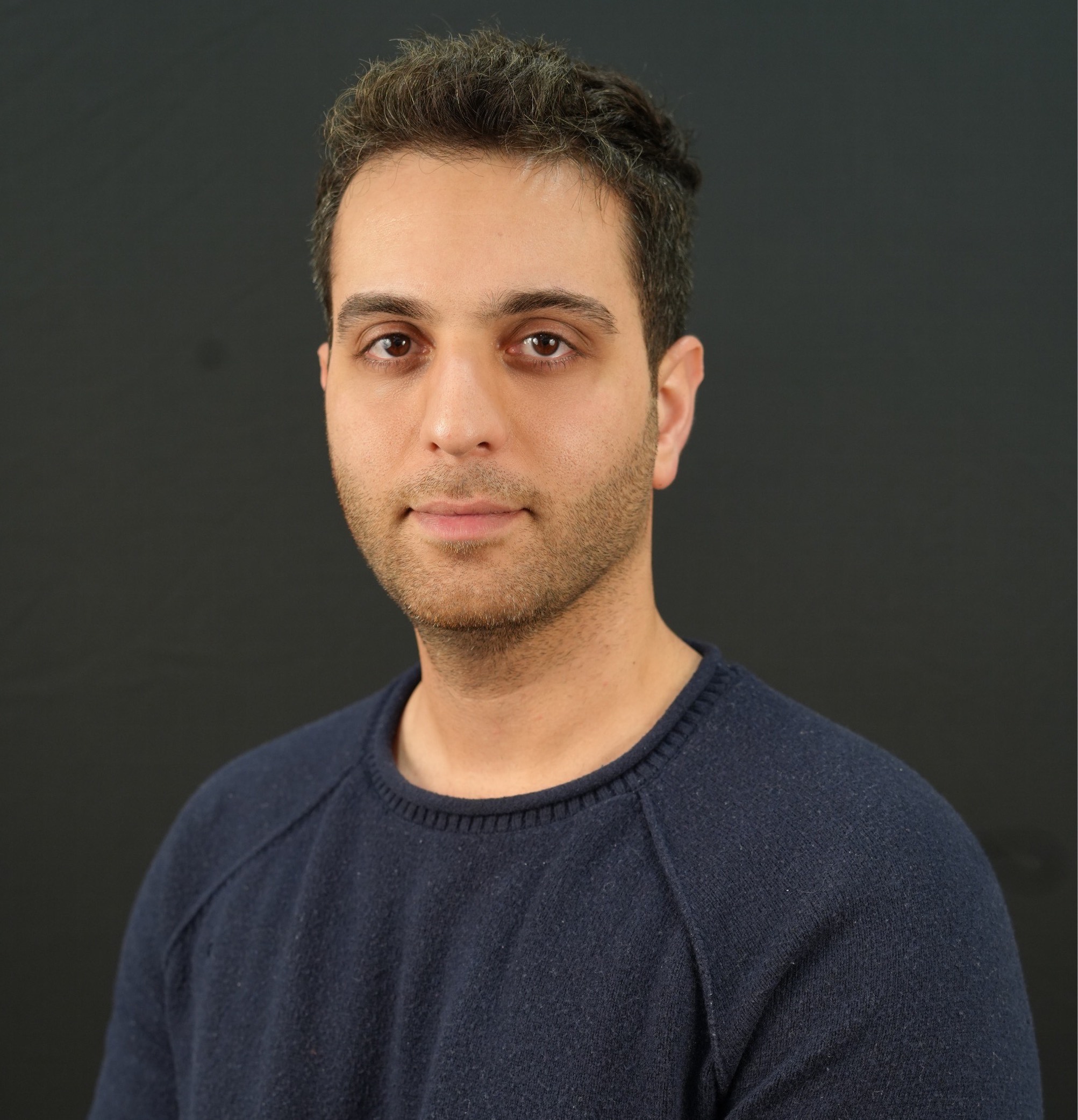}}]{Erfaun Noorani} earned his Ph.D. and M.S. in Electrical and Computer Engineering from the University of Maryland, College Park, where he was a Clark Doctoral Fellow. He also holds a B.Sc. degree in Electrical Engineering from Drexel University, Philadelphia, PA. He is currently a Technical Staff at MIT Lincoln Laboratory. Prior to this role, Erfaun was a Postdoctoral Associate within the Institute for Systems Research (ISR) at the University of Maryland, College Park. Erfaun's research focuses on robust and risk-sensitive reinforcement learning.
\end{IEEEbiography}

\begin{IEEEbiography}[{\includegraphics[width=1in,height=1.25in,clip,keepaspectratio]
{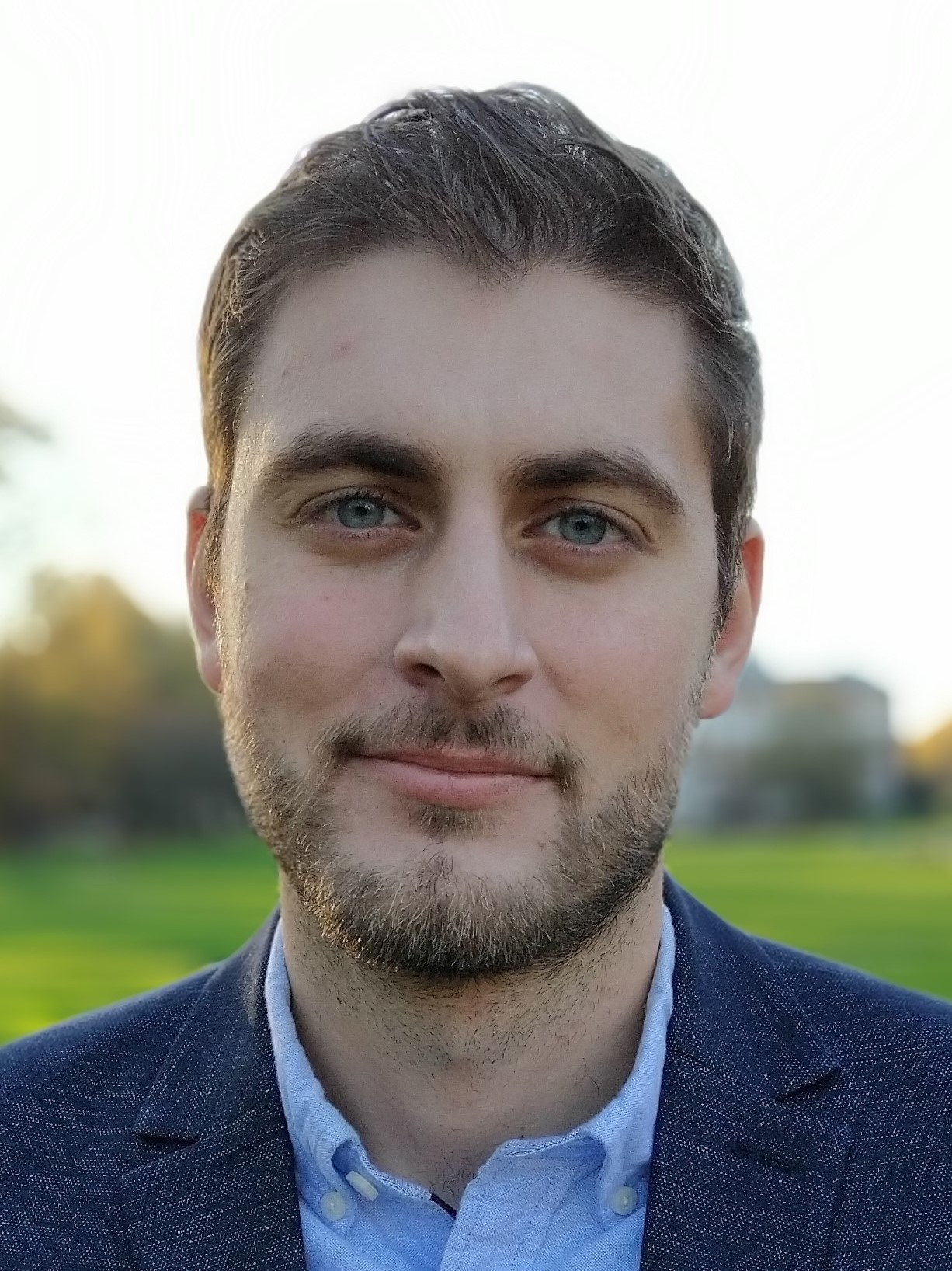}}]{Christos N. Mavridis} (M'20) 
received his Diploma in electrical and computer engineering from the National Technical University of Athens, Greece, in 2017,
and the M.S. and  Ph.D. degrees in electrical and computer engineering at the University of Maryland, College Park, MD, in 2021. 
His research interests include hybrid systems and control theory, stochastic optimization, and learning theory.

He is currently a postdoc at KTH Royal Institute of Technology, Stockholm, and has been affiliated as a research scientist with the Institute for Systems Research (ISR), University of Maryland, MD, the Nokia Bell Labs, NJ, the Xerox Palo Alto Research Center (PARC), CA, and Ericsson AB, Stockholm. 

Dr. Mavridis is an IEEE member, and a member of IEEE/CSS Technical Committee on Security and Privacy. He has received the A. James Clark School of Engineering Distinguished Graduate Fellowship and the Ann G. Wylie Dissertation Fellowship in 2017 and 2021, respectively. He has been a finalist in the Qualcomm Innovation Fellowship US, San Diego, CA, 2018, and he has received the Best Student Paper Award in the IEEE International Conference on Intelligent Transportation Systems (ITSC), 2021.
\end{IEEEbiography}


\begin{IEEEbiography}[{\includegraphics[width=1in,height=1.25in,clip,keepaspectratio]
{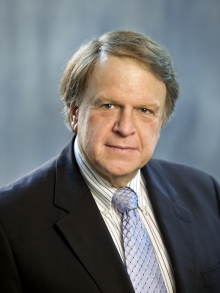}}]{John S. Baras} (LF'13) 
received the Diploma degree in electrical and mechanical engineering from the National Technical University of Athens, Greece, in 1970, and the M.S. and Ph.D. degrees in applied mathematics from Harvard University, Cambridge, MA, USA, in 1971 and 1973, respectively.

He is a Distinguished University Professor and holds the Lockheed Martin Chair in Systems Engineering, with the Department of Electrical and Computer Engineering and the Institute for Systems Research (ISR), at the University of Maryland College Park. From 1985 to 1991, he was the Founding Director of the ISR. Since 1992, he has been the Director of the Maryland Center for Hybrid Networks (HYNET), which he co-founded. His research interests include systems and control, optimization, communication networks, applied mathematics, machine learning, artificial intelligence, signal processing, robotics, computing systems, security, trust, systems biology, healthcare systems, model-based systems engineering.

Dr. Baras is a Fellow of IEEE (Life), SIAM, AAAS, NAI, IFAC, AMS, AIAA, Member of the National Academy of Inventors and a Foreign Member of the Royal Swedish Academy of Engineering Sciences. Major honors include the 1980 George Axelby Award from the IEEE Control Systems Society, the 2006 Leonard Abraham Prize from the IEEE Communications Society, the 2017 IEEE Simon Ramo Medal, the 2017 AACC Richard E. Bellman Control Heritage Award, the 2018 AIAA Aerospace Communications Award. In 2016 he was inducted in the A. J. Clark School of Engineering Innovation Hall of Fame. In 2018 he was awarded a Doctorate Honoris Causa by his alma mater the National Technical University of Athens, Greece.   
\end{IEEEbiography}

\end{document}